\documentclass[preprint,tighten,nofootinbib,aps,amssymb,amsmath,floatfix]{revtex4}
\usepackage{psfig,epsfig,colordvi,amssymb,amsmath,bbm,graphics}
\usepackage{lscape}
\usepackage{rotfloat}
\usepackage{rotating}

\def\lsim{\mathrel{\rlap{\lower4pt\hbox{\hskip1pt$\sim$}}
    \raise1pt\hbox{$<$}}}                
\def\gsim{\mathrel{\rlap{\lower4pt\hbox{\hskip1pt$\sim$}}
    \raise1pt\hbox{$>$}}}                

\newcommand{\be}{\begin{equation}}
\newcommand{\ee}{\end{equation}}
\newcommand{\bea}{\begin{eqnarray}} 
\newcommand{\eea}{\end{eqnarray}}

\newcommand{\csw}{\, c_{\rm SW}}

\begin{document}
\setlength{\textheight}{23cm}
\textheight=23cm
\draft

\vspace*{-25truemm}
\begin{flushright}
\hspace*{-1cm}ROM2F/2010/20, RM3-TH/10-11
 
\end{flushright}\vspace{5truemm}

\title{Perturbative renormalization factors and ${\cal O}(a^2)$
  corrections for lattice 4-fermion operators with improved
  fermion/gluon actions 
\begin{figure}[h]
  \begin{center}
    \includegraphics[scale=0.17]{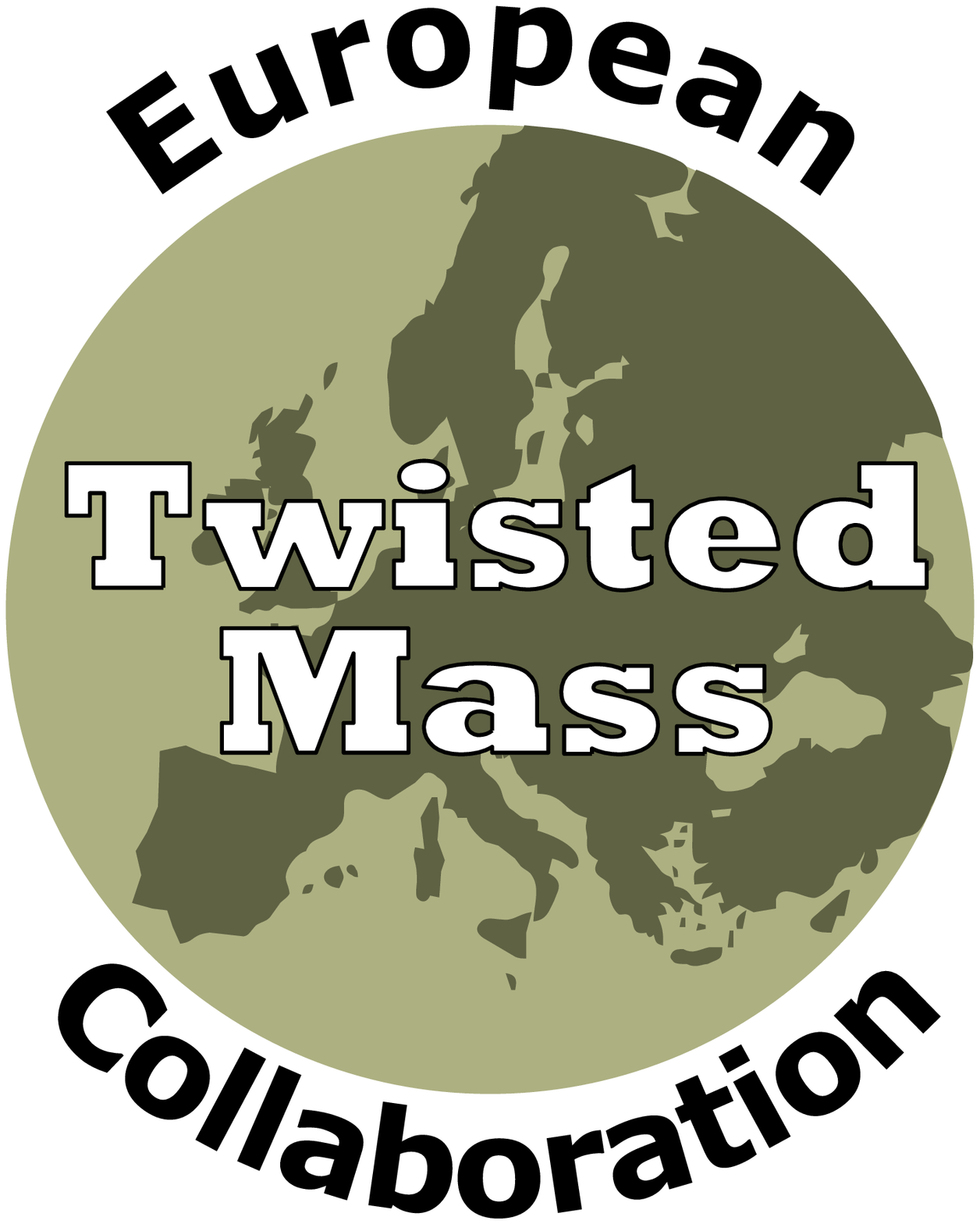}
  \end{center}
\end{figure}}

\vspace{-0.5cm}
\author{Martha Constantinou$^a$, Petros Dimopoulos$^b$, Roberto Frezzotti$^{cd}$,
  Vittorio Lubicz$^{ef}$, Haralambos Panagopoulos$^a$, Apostolos
  Skouroupathis$^a$, Fotos Stylianou$^a$ 
\vspace{0.5cm}
\vskip 0 true mm
{\it $^a$ Department of Physics, University of Cyprus}
{\it P.O. Box 20537, Nicosia CY-1678, Cyprus}
\vskip 1 true mm
{\it $^b$ Dipartimento di Fisica, Sapienza, Universit\`a di Roma}
{\it Piazzale A. Moro, I-00185 Rome, Italy}
\vskip 1 true mm
{\it $^c$ Dipartimento di Fisica, Universit\`a di Roma ``Tor Vergata''}
{\it Via della Ricerca Scientifica 1, I-00133 Rome, Italy}
\vskip 1 true mm
{\it $^d$ INFN, Sezione di ``Tor Vergata"}
{\it c/o Dipartimento di Fisica, Universit\`a di Roma ``Tor Vergata''}
{\it Via della Ricerca Scientifica 1, I-00133 Rome, Italy}
\vskip 1 true mm
{\it $^e$ Dipartimento di Fisica, Universit\`a Roma Tre}
{\it Via della Vasca Navale 84, I-00146 Rome, Italy}
\vskip 1 true mm
{\it $^f$ INFN, Sezione di Roma Tre}
{\it c/o Dipartimento di Fisica, Universit\`a Roma Tre}
{\it Via della Vasca Navale 84, I-00146 Rome, Italy}\\[2ex]
 E-mail: marthac@ucy.ac.cy, petros.dimopoulos@roma2.infn.it,
 roberto.frezzotti@roma2.infn.it, lubicz@fis.uniroma3.it,
         haris@ucy.ac.cy, php4as01@ucy.ac.cy, fstyli01@ucy.ac.cy}

\maketitle

\centerline{Abstract}

  In this work we calculate the corrections to the amputated Green's
  functions of 4-fermion operators, in 1-loop Lattice Perturbation
  theory. One of the novel aspects of our
  calculations is that they are carried out to second order in the
  lattice spacing, ${\cal O}(a^2)$.

  We employ the Wilson/clover action for massless fermions (also
  applicable for the twisted mass action in the chiral limit) and a
  family of Symanzik improved actions for gluons. Our calculations have been
  carried out in a general covariant gauge. Results have been obtained
  for several popular choices of values for the Symanzik coefficients
  (Plaquette, Tree-level Symanzik, Iwasaki, TILW and DBW2 action).

  While our Green's function calculations regard any pointlike
  4-fermion operators which do not mix with lower dimension ones,
  we pay particular attention to $\Delta F=2$
  operators, both Parity Conserving and Parity Violating ($F$ stands
  for flavour: $S,\,C,\,B$). By appropriately projecting those bare Green's
  functions we compute the perturbative renormalization constants for
  a complete basis of 4-fermion operators and we study their mixing pattern.
  For some of the actions considered here, even ${\cal O}(a^0)$ results
  did not exist in the literature to date. The correction terms which
  we calculate (along with our previous ${\cal O}(a^2)$ calculation of
  $Z_\Psi$~\cite{CLPS,ETMC,ACKPS}) are essential ingredients for minimizing
  the lattice artifacts which are present in non-perturbative
  evaluations of renormalization constants with the RI$'$-MOM method.

   Our perturbative results, for the matrix elements of $\Delta F=2$
   operators and for the corresponding renormalization matrices,
   depend on a large number of parameters: coupling constant, number
   of colors, lattice spacing, external momentum, clover parameter,
   Symanzik coefficients, gauge parameter. To make these results most
   easily accessible to the reader, we have included them in the
   distribution package of this paper, as an ASCII file named:
   4-fermi.m; the file is best perused as Mathematica input.
 
   The main results of this work have been applied to improve
       non-perturbative estimates of the $B_K$-parameter in $N_F=2$
       twisted mass lattice QCD~\cite{ETMC_BK}.

\section{Introduction}
A number of flavour-changing processes are currently under study in
Lattice simulations. Among the most common examples are the
decay $K \rightarrow \pi \pi$ and $K^0$--$\bar K^0$ oscillations. 
From experimental evidence, we know that these
weak processes violate the CP symmetry. In theory, the calculation of
the amount of CP violation in $K^0$--$\bar K^0$ oscillations
requires the knowledge of the kaon $B_K$ parameter.

The parameter $B_K$ is obtained from the $\Delta S = 2$ weak
matrix element:
\be
B_K = \frac{\langle \bar K^0|\hat O^{\Delta S=2}| K^0 \rangle}
           {\frac{8}{3}\langle \bar K^0|\bar s \gamma_\mu d|0 \rangle \, \langle 0|\bar s \gamma_\mu d|K^0 \rangle} \, ,
\ee
where $s$ and $d$ stand for strange and down quarks, and $\hat
O^{\Delta S=2}$ is the effective 4-quark interaction renormalized
operator, corresponding to the bare operator:
\be
O^{\Delta S=2} = (\bar s \gamma_\mu^{\rm L} d)(\bar s \gamma_\mu^{\rm L} d) ,
\qquad \gamma_\mu^{\rm L}=\gamma_\mu(\openone-\gamma_5).
\label{O_bare}
\ee
The above operator splits into parity-even and parity-odd parts; in
standard notation:
$O^{\Delta S=2} = O_{VV+AA}^{\Delta S=2} - O_{VA+AV}^{\Delta S=2}$. 
Since the above weak process is
simulated in the framework of Lattice QCD, where Parity is a symmetry,
the parity-odd part gives no contribution to the $K^0$--$\bar K^0$
matrix element. Thus, we conclude
that $B_K$ can be extracted from the correlator ($x_0\!>\!0$,
$y_0\!<\!0$): 
{\small{
\be 
C_{KOK}(x,y) = \langle (\bar{d}\gamma_5 s)(x) \hat{O}_{VV+AA}^{\Delta S=2}(0) (\bar{d}\gamma_5 s)(y) \rangle, \qquad
O_{VV+AA}^{\Delta S=2} = (\bar{s} \gamma_\mu d) (\bar{s}
\gamma_\mu d) + (\bar{s} \gamma_\mu\gamma_5 d) (\bar{s}
\gamma_\mu\gamma_5 d) \, ,
\label{VVAA}
\ee
}}
\hspace{-0.15cm}where $O_{VV+AA}^{\Delta S=2}$ is the bare operator and
$\hat{O}_{VV+AA}^{\Delta S=2}$ is the corresponding renormalized
operator.

Our results are immediately applicable to other $\Delta F=2$ processes
of great phenomenological interest, such as $D-\bar D$ or $B-\bar B$
mixing. They are also useful in new physics models (i.e. beyond the
standard model), because there the complete basis of 4-fermion
operators contributes to neutral meson mixing amplitudes; this is the
case for instance of SUSY models (see e.g. \cite{SUSY}). For this, one
needs to study more general operators of the form 
\be
{\cal O}_{XY} \equiv (\bar{s}\,X\,d)(\bar{s}\,Y\,d)
\label{O_basis}
\ee
where $X$ and $Y$ are general Dirac matrices (see Eq.~(\ref{Gamma})).

With Wilson fermions on the lattice, the explicit breaking of
chiral symmetry also induces a mixing between the Standard Model
operator in Eq.~(\ref{O_bare}) and the other $\Delta S=2$ operators of the
basis Eq.~(\ref{O_basis}). A strategy which allows to avoid this mixing and, at
the same time, guarantees automatic ${\cal O}(a)$-improvement of
the four fermion operators has been proposed in \cite{FR} and it makes
use of twisted and Osterwalder-Seiler fermions. In this approach,
for the $B_K$ computation, in place of the operator in Eq.~(\ref{VVAA}) a
four-quark operator with a different flavour content ($s$, $d$, $s'$,
$d'$), and with $\Delta S = \Delta s +\Delta s'=2$ is
considered, namely \cite{FR}
\be
 {\cal O}_{VV+AA}^{\Delta S=2} =
(\bar{s} \gamma_\mu d) (\bar{s}' \gamma_\mu d')
+ (\bar{s} \gamma_\mu\gamma_5 d) (\bar{s}' \gamma_\mu\gamma_5 d')
+ (\bar{s} \gamma_\mu d') (\bar{s}' \gamma_\mu d)
+ (\bar{s} \gamma_\mu\gamma_5 d') (\bar{s}' \gamma_\mu\gamma_5 d) \, ,
\label{VVAAprime}
\ee
where now the correlator is given by:
%
$C_{K{\cal O}K'}(x,y) = \langle
(\bar{d}\gamma_5 s)(x) \, 2 {\cal O}_{VV+AA}^{\Delta S=2}(0)
(\bar{d}'\gamma_5 s')(y) \rangle $. 
%
Making use of Wick's theorem one checks the equality:
$C_{K{\cal O}K'}(x,y) = C_{KOK}(x,y)$,
which means that both correlators contain the same physical
information.

The aforementioned matrix elements are very sensitive to various systematic
errors. A major issue facing Lattice Gauge Theory, since its early
days, has been the reduction of effects induced by the finiteness of
lattice spacing $a$, in order to better approach the elusive continuum
limit. 

In order to obtain reliable non-perturbative estimates of physical
quantities it is essential to
keep under control the ${\cal O}(a)$ systematic errors in simulations
or, additionally, reduce the lattice artifacts in numerical results. 
Such a reduction, regarding renormalization functions, can be achieved by
subtracting appropriately the ${\cal O}(a^2)$ perturbative correction terms
presented in this paper, from corresponding non-perturbative results.

In this paper we address the perturbative aspects of this
problem from a very general point of view. In particular, we study the
bare 4-point amputated Green's function of the most general pointlike
4-fermion operators with four distinct flavours\footnote{For
$\Delta S=1$ operators with flavour structure $(\bar s X d)(\bar q Y
q)$ penguin contractions induce a power divergent mixing of the four
fermion operators with lower dimension operators. This case is not
considered in the present paper.}. Although the computational
procedure laid out in the paper is applicable to all orders in the
lattice spacing, we focus on two different results: \\
{\bf{1.}} The perturbative 1-loop evaluation of renormalization
factors for a variety of 4-fermion operators. These factors can be
used to renormalize 4-fermion operators computed non-perturbatively
with any fermion/gluon Wilson-like improved action. This part can be
considered as an extension of other computations of 4-fermion operator
renormalization~\cite{Aoki}.\\
{\bf{2.}} The evaluation of ${\cal O}(a^2)$ contributions to
the aforementioned 1-loop computations. These are very useful to
improve non-perturbative estimates for the same renormalization
factors~\cite{ETMC_BK}, since they can reduce lattice artifacts,
leading to more reliable determinations.

In Section~\ref{Sec_oper} we define the general 4-fermion operators
and describe the setup of the computation. The calculations are
carried out up to 1-loop in Lattice Perturbation theory and up to
${\cal O}(a^2)$ in lattice spacing. In the same Section we also
present simplified expressions for the three Feynman diagrams, which
constitute the building blocks of the whole calculation. 
In addition, we address certain difficulties which are associated to
the ${\cal O}(a^2)$ computation. In Section \ref{Sec_renorm} we switch
to the evaluation of the renormalization
matrices for the 4-fermion operators. In particular, we focus on the
complete basis of 20 four-fermion operators of dimension six which
do not need power subtractions (i.e. mixing occurs only with other
operators of equal dimensions). In the last Section, we summarize the
main results of this work, and discuss how non-perturbative estimates
are being improved by subtracting our ${\cal O}(a^2)$ correction terms.

\section{Amputated Green's functions of 4-fermion $\Delta S= \Delta s + \Delta s'=2$ operators.}
\label{Sec_oper}

Here we evaluate, up to ${\cal O}(a^2)$, the 1-loop matrix
element of the 4-fermion operators (the superscript letter F
  stands for Fierz.):
{\small{
\bea
{\cal O}_{XY} \equiv (\bar{s}\,X\,d)(\bar{s}'\,Y\,d')\equiv 
\sum_x \sum_{c,d} \sum_{k_1,\,k_2,\,k_3,\,k_4}
\Bigl(\bar{s}_{{k_1}}^c(x)  \,X_{{k_1 k_2}}\,  d_{{k_2}}^c(x)\Bigr) 
\Bigl(\bar{s'}_{{k_3}}^d(x) \,Y_{{k_3 k_4}}\, {d'}_{{k_4}}^d(x)\Bigr) \\
\label{O_XY}
{\cal O}^F_{XY} \equiv (\bar{s}\,X\,d')(\bar{s}'\,Y\,d)\equiv 
\sum_x \sum_{c,d} \sum_{k_1,\,k_2,\,k_3,\,k_4}
\Bigl(\bar{s}_{{k_1}}^c(x)  \,X_{{k_1 k_2}}\, {d'}_{{k_2}}^c(x)\Bigr) 
\Bigl(\bar{s'}_{{k_3}}^d(x) \,Y_{{k_3 k_4}}\, d_{{k_4}}^d(x)\Bigr)
\label{O^F_XY}
\eea
}}
\hspace{-0.15cm}with a generic initial state: 
$\bar{d'}_{{i_4}}^{{a_4}}(p_4)\,{s'}_{{i_3}}^{{a_3}}(p_3) |0\rangle $,
and a generic final state:
$\langle 0 | \bar{d}_{{i_2}}^{{a_2}}(p_2)\,s_{{i_1}}^{{a_1}}(p_1)$.
Spin indices are denoted by $i,\,k$, and color indices by
$a,\,c,\,d$, while  
$X$ and $Y$ correspond to the following set of products of the Dirac
matrices:
{\small{
\be
X,\,Y = \{\openone,\, \gamma^5,\, \gamma_\mu,\, \gamma_\mu \gamma^5,
  \,\sigma_{\mu\nu}, \,\gamma^5\sigma_{\mu\nu}\} \equiv
  \{S,P,V,A,T,\tilde T \}; 
  \qquad \sigma_{\mu\nu}=\frac{1}{2}[\gamma_\mu,\gamma_\nu].
\label{Gamma}
\ee
}}
\hspace{0.65cm}Our calculations are performed using massless fermions described by
the Wilson/clover action. By taking $m_f=0$, our results 
are identical also for the twisted mass and the
Osterwalder-Seiler actions in the chiral limit (in the so called
twisted mass basis). For gluons we employ a 3-parameter family of
Symanzik improved actions, which comprises all common gluon actions
(Plaquette, tree-level Symanzik, Iwasaki, DBW2, L\"uscher-Weisz). 
Conventions and notations for the actions, as well as algebraic
manipulations involving the evaluation of 1-loop Feynman diagrams (up
to ${\cal O}(a^2)$), are described in detail in Ref. \cite{CLPS}.

To establish notation and normalization, let us first write the
tree-level expression for the amputated Green's functions of the
operators ${\cal O}_{XY}$ and ${\cal O}^F_{XY}$:
{\small{
\be
\Lambda^{XY}_{tree}(p_1,p_2,p_3,p_4,r_s,r_d,r_{s'},r_{d'})_
{{i_1 i_2 i_3 i_4}}^{{a_1 a_2 a_3 a_4}} = 
X_{i_1 i_2}\, Y_{i_3 i_4}\, \delta_{a_1 a_2}\,\delta_{a_3 a_4} ,
\label{L_XY_tree}
\ee
\be
(\Lambda^F)^{XY}_{tree}(p_1,p_2,p_3,p_4,r_s,r_d,r_{s'},r_{d'})_
{{i_1 i_2 i_3 i_4}}^{{a_1 a_2 a_3 a_4}} = 
- X_{i_1 i_4}\, Y_{i_3 i_2}\, \delta_{a_1 a_4}\,\delta_{a_3 a_2} ,
\ee
}}
\hspace{-0.15cm}where $r$ is the Wilson parameter, one for each flavour. 

\begin{figure}
\includegraphics[height=3.20truecm]{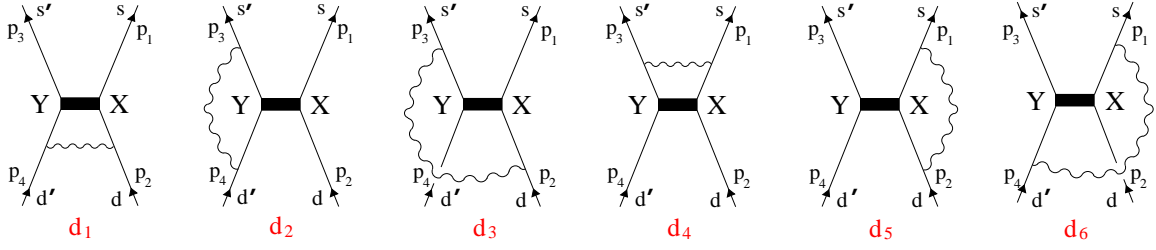}
\hskip -0.5cm
\caption{1-loop diagrams contributing to the amputated Green's
function of the 4-fermion operator ${\cal O}^{XY}$. Wavy (solid) lines
represent gluons (fermions).}
\label{fig1}
\end{figure}

We continue with the first quantum corrections. There are twelve 1-loop
diagrams that enter our 4-fermion calculation, six for each operator
${\cal O}_{XY}$, ${\cal O}^F_{XY}$. The diagrams $d_1-d_6$
corresponding to the operator ${\cal O}_{XY}$ are illustrated in
Fig. \ref{fig1}. The other six diagrams,
$d^F_1-d^F_6$, involved in the Green's function of ${\cal O}^F_{XY}$ 
are similar to $d_1-d_6$, and may be obtained from $d_1-d_6$ by
interchanging the fermionic fields $d$ and $d'$ along with their
momenta, color and spin indices, and respective Wilson parameters. 

The only diagrams that need to be calculated from first principles are
$d_1$, $d_2$ and $d_3$, while the rest can be expressed in terms of the
first three. This is a result of the symmetries between the diagrams
($d_1,\,d_4$), ($d_2,\,d_5$) and ($d_3,\,d_6$). Diagrams $d_4$, $d_5$
and $d_6$ can be expressed as diagrams $d_1$, $d_2$ and $d_3$ by
exchanging the external quark legs and $X,\,Y$, if necessary. In
particular, the expressions for the amputated Green's functions
$\Lambda_{d_4}^{XY}-\Lambda_{d_6}^{XY}$ can be obtained via the
following relations: 
{\small{
\hspace{-0.5cm}
\bea
\hspace{-0.5cm}
\Lambda_{d_4}^{XY}(p_1,p_2,p_3,p_4,r_s,r_d,r_{s'},r_{d'})_{i_1 i_2 i_3 i_4}^{a_1 a_2 a_3 a_4} &=&
\left(\Lambda_{d_1}^{XY}({-}p_2,{-}p_1,{-}p_4,{-}p_3,r_d,r_s,r_{d'},r_{s'})_{i_2 i_1 i_4 i_3}^{a_2 a_1 a_4 a_3}\right)^\star \label{Lambda_4},\,\,\\ 
\Lambda_{d_5}^{XY}(p_1,p_2,p_3,p_4,r_s,r_d,r_{s'},r_{d'})_{i_1 i_2 i_3 i_4}^{a_1 a_2 a_3 a_4} &=&\phantom{\Bigl(}
\Lambda_{d_2}^{YX}      (p_3,p_4,p_1,p_2,r_{s'},r_{d'},r_s,r_d)_{i_3 i_4 i_1 i_2}^{a_3 a_4 a_1 a_2},\\ 
\Lambda_{d_6}^{XY}(p_1,p_2,p_3,p_4,r_s,r_d,r_{s'},r_{d'})_{i_1 i_2 i_3 i_4}^{a_1 a_2 a_3 a_4} &=&\phantom{\Bigl(}
\Lambda_{d_3}^{YX} (p_3,p_4,p_1,p_2,r_{s'},r_{d'},r_s,r_d)_{i_3 i_4 i_1 i_2}^{a_3 a_4 a_1 a_2}.
\eea
}}
\hspace{-0.15cm} Once we have constructed
$\Lambda_{d_4}^{XY}-\Lambda_{d_6}^{XY}$ we can use relation:
{\small{
\be
(\Lambda^F)_{d_j}^{XY}(p_1,p_2,p_3,p_4,r_s,r_d,r_{s'},r_{d'})_{i_1 i_2 i_3 i_4}^{a_1 a_2 a_3 a_4} =
-\Lambda_{d_j}^{XY}      (p_1,p_4,p_3,p_2,r_s,r_{d'},r_{s'},r_d)_{i_1 i_4 i_3 i_2}^{a_1 a_4 a_3 a_2}, 
\ee
}}
\hspace{-0.15cm}to derive the expressions for $(\Lambda^F)_{d_i}^{XY}$ ($i=1,\cdots, 6$).
From the amputated Green's functions for all
twelve diagrams we can write down the total 1-loop
expressions for the operators ${\cal O}_{XY}$ and ${\cal O}^F_{XY}$:
{\small{
\be
\Lambda_{1-loop}^{XY} = \sum_{j=1}^{6}\Lambda_{d_j}^{XY},\qquad
(\Lambda^F)_{1-loop}^{XY}= \sum_{j=1}^{6}(\Lambda^F)_{d_j}^{XY}.
\label{L^F_XY_1loop}
\ee
}}
In our algebraic expressions for the 1-loop amputated Green's
functions $\Lambda_{d_1}^{XY}$, $\Lambda_{d_2}^{XY}$ and
$\Lambda_{d_3}^{XY}$ we kept the Wilson parameters for each quark
field distinct, that is: $\,r_s$, $r_d$, $r_{s'}$, $r_{d'}$ for the
quark fields $s$, $d$, $s'$ and $d'$ respectively. For the required
numerical integration of the algebraic expressions corresponding to
each Feynman diagram, we are forced to choose the value for each $r$
parameter.In the numerical results presented in this paper we
set:
\vspace{-0.25cm}
\be
r_s=r_d=r_{s'}=r_{d'}=1. 
\ee
\vskip -0.25cm
Concerning the external momenta $p_i$ (shown explicitly in
Fig. \ref{fig1}) we have chosen to evaluate the
amputated Green's functions at the renormalization point:
\vspace{-0.25cm}
\be
p_1 = p_2 = p_3 = p_4 \equiv p.
\ee
\vskip -0.25cm
It is easy and not time consuming to repeat the calculations for other
choices of Wilson parameters and for other renormalization prescriptions.
The final 1-loop expressions for $\Lambda_{d_1}^{XY}$, $\Lambda_{d_2}^{XY}$
and $\Lambda_{d_3}^{XY}$, up to ${\cal O}(a^2)$, are obtained as a
function of: the coupling constant $g$, clover
parameter $\csw$, number of colors $N_c$, lattice spacing $a$, external
momentum $p$ and gauge parameter $\lambda$.

As an example we present the results for
$\Lambda_{d_1}^{XY}$ and for the special choices: 
$\csw=0$, $\lambda=0$ (Landau Gauge), $r_s=r_d=r_{s'}=r_{d'}=1$, and
tree-level Symanzik improved action: 
%
{\footnotesize{
\bea
\hspace{-1.25cm}
&&\Blue{\Lambda_{d_1}^{XY}(p)_{i_1\,i_2\,i_3\,i_4}^{a_1\,a_2\,a_3\,a_4}}=\frac{g^2}{16\pi^2}
\left(\delta_{a_1\,a_4}\delta_{a_3\,a_2}-\frac{\delta_{a_1\,a_2}\delta_{a_3\,a_4}}{N_c}\right)
{\times}
\bigg \{
\Blue{(\Lambda_{{\cal O}(a^0)})^{XY}_{d_1}}+\Red{a}\Blue{(\Lambda_{{\cal O}(a^1)})^{XY}_{d_1}}+\Red{a^2}\Blue{(\Lambda_{{\cal O}(a^2)})^{XY}_{d_1}}
\bigg \} \,,
\label{Lambda_d1}
\eea }}
where:
{\footnotesize{
\vspace{-0.2cm}
\bea
\hspace{-1.0cm} &&\Blue{(\Lambda_{{\cal O}(a^0)})^{XY}_{d_1}} = 
X_{i_1\,i_2}Y_{i_3\,i_4}\left[-\frac{1}{2}\ln(\Red{a^2}p^2)-0.05294144(3)\right]
+\sum_\mu (X\gamma^\mu)_{i_1\,i_2}(Y\gamma^\mu)_{i_3\,i_4}\left[-0.507914049(6)\right] \nonumber \\
&&+ \sum_{\mu,\nu}
(X\gamma^\mu\gamma^\nu)_{i_1\,i_2}(Y\gamma^\mu\gamma^\nu)_{i_3\,i_4}\left[\frac{1}{8}\ln(\Red{a^2}p^2)+0.018598520(2)\right]\nonumber \\
&&+ \sum_{\mu,\nu,\rho} (X\gamma^\mu\gamma^\rho)_{i_1\,i_2}(Y\gamma^\nu\gamma^\rho)_{i_3\,i_4}\left[0.397715726853\frac{p_\mu p_\nu}{p^2}\right]\,.
\eea
}}
The ${\cal O}(a^1)$ and ${\cal O}(a^2)$  contributions of
Eq.~(\ref{Lambda_d1}) along with the complete results for all diagrams
for tree-level Symanzik improved gluons, $\csw\neq 0$ and $\lambda\neq 0$,
are presented in Appendix A; the reader can find similar expressions
for other gluon actions in electronic form (4-fermi.m). We note in
passing that in diagram 3 the dependence on external momentum has the
same terms as in diagram 2, with identical numerical coefficients; the
difference between the two diagrams lies in the structure of color and
gamma matrices multiplying each term.  

The setup presented up to this point applies to both ${\cal O}(a^0)$
and ${\cal O}(a^2)$ calculation. For the ${\cal O}(a^2)$ case
additional difficulties arise in extracting correctly the full ${\cal
  O}(a^2)$ dependence.
The crucial point of our calculation is the correct extraction of the
full ${\cal O}(a^2)$ dependence from loop integrands with strong IR
divergences (convergent only beyond 6 dimensions). The singularities
are isolated using the procedure explained in
Ref. \cite{CLPS}. In order to reduce the number of strong IR
divergent integrals, appearing in diagram $d_1$, we have inserted the
identity below into selected 3-point functions: 
{\small{
\be
1=\frac{1}{\widehat{a\,p}^2}\Bigl(\widehat{k+a\,p}^2 +
\widehat{k-a\,p}^2 - 2 \hat k^2 + 16 \sum_\sigma \sin(k_\sigma)^2
\sin(a p_\sigma)^2\Bigr),
\label{identity} 
\ee
}}
\hspace{-0.15cm}where $\hat{q}^2=4\sum_\mu \sin^2(\frac{q_\mu}{2})$ and $k\,(p)$ is
the loop (external) momentum. Repeated use of Eq. (\ref{identity})
reduces the 3-point functions to either 2-point functions or more
convergent expressions. The factor $1/\widehat{a\,p}^2$ in Eq. (\ref{identity})
can be treated by Taylor expansion. For our calculations it was
necessary only to ${\cal O}(a^0)$: 
{\small{
\be
\frac{1}{\widehat{a\,p}^2} =
\frac{1}{a^2\,p^2} + \frac{\sum_\sigma p_\sigma^4}{(p^2)^2} +{\cal O}(a^2\,p^2) .
\ee
}}
\hspace{-0.15cm}Here we present one of the four integrals
with strong IR divergences that enter this calculation:
{\footnotesize{
\bea
\Blue{\int_{-\pi}^\pi \frac{d^4k}{(2\pi)^4}\frac{\sin(k_\mu)\,\sin(k_\nu)}{\hat k^2\, 
\widehat{k+a\,p}^2\, \widehat{k-a\,p}^2}} &=& 
\delta_{\mu\nu}\Bigl(0.002457072288 -\frac{\ln(\Red{a^2}p^2)}{64\pi^2}\Bigr)
+0.001870841540 \frac{p_\mu\,p_\nu}{p^2}\nonumber\\
&+& \Red{a^2}\Biggl[
\delta_{\mu\nu} \Biggl(\,p^2 \Bigl( 0.00055270353(6) - \frac{\ln(\Red{a^2}p^2)}{512\pi^2}\Bigr)\nonumber\\
&&\phantom{a^2}- p_\mu^2 \Bigl(0.0001282022(1) +\frac{\ln(\Red{a^2}\,p^2)}{768\pi^2}\Bigr)
+0.000157122310\,\frac{\sum_\sigma p_\sigma^4}{p^2}\Biggl) \nonumber \\
&&\phantom{a^2}+ p_\mu\,p_\nu \Biggl(-0.00029731225(4)+\frac{\ln(\Red{a^2}p^2)}{768\pi^2} 
-0.000047949674 \frac{(p_\mu^2+p_\nu^2)}{p^2} \nonumber \\
&&\phantom{a^2}+ 0.000268598599\frac{\sum_\sigma p_\sigma^4}{(p^2)^2} \Biggr)\Biggl] +{\cal O}(\Red{a^4}\,p^4).
\eea
}} 
\noindent The results for the other three integrals can be found in Ref. \cite{CLPS}.
Integrands with simple IR divergences (convergent beyond 4 dimensions)
can be handled by well-known techniques. 

\section{Mixing and Renormalization of ${\cal O}_{XY}$ and ${\cal O}^F_{XY}$ on the lattice.}
\label{Sec_renorm}
The matrix element $\langle \bar K^0|O^{\Delta S=2}_{VV+AA}| K^0 \rangle$ 
is very sensitive to various systematic errors. The main roots of this
problem are: 
{\bf a)} ${\cal O}(a)$ systematic errors due to numerical integration,
{\bf b)} with Wilson-like fermions, the operator $O^{\Delta
  S=2}_{VV+AA}$ mixes with other 4-fermion $\Delta S=2$ operators of
dimension six. Mixing with operators of lower dimensionality is
impossible because there is no candidate $\Delta S=2$ operator.

In order to address these problems we have calculated the mixing
pattern (renormalization matrices) of the Parity Conserving and Parity
Violating 4-fermion $\Delta S=2$ operators (defined below), by using
the amputated Green's functions obtained in the previous
section. A more extensive theoretical background and non-perturbative
results, concerning renormalization matrices of 4-fermion operators,
can be found in Ref. \cite{DGMTV} (see also
\cite{FR,BGLMPR,MPSTV}). Next we summarize all important relations
from Ref. \cite{DGMTV} needed for the present calculation.

One can construct a complete basis of 20
independent operators which have the symmetries of the generic QCD Wilson lattice
action (Parity $P$, Charge
conjugation $C$, Flavour exchange symmetry $S {\equiv} (d
\leftrightarrow d')$, Flavour Switching symmetries $S' {\equiv}
(s \leftrightarrow d , s' \leftrightarrow d')$ and
$S'' {\equiv} (s \leftrightarrow d' , d \leftrightarrow s')$), with 4 degenerate quarks.
This basis can be decomposed into smaller independent bases according
to the discrete symmetries $P,\,S,\,CPS',\,CPS''$. Following the notation
of Ref. \cite{DGMTV} we have 10 Parity Conserving operators, $Q$,
($P{=}+1,\,S{=}\pm 1$)
and 10 Parity Violating operators, $\cal Q$, ($P{=}-1,\,S{=}\pm 1$):

\noindent
{\small{
\begin{minipage}{8cm}
\bea
\begin{cases}
Q_1^{S=\pm 1}\equiv \frac{1}{2}\left[{\cal O}_{VV} \pm {\cal O}^F_{VV}\right]+\frac{1}{2}\left[{\cal O}_{AA} \pm {\cal O}^F_{AA}\right],\nonumber\\[0.4ex]
Q_2^{S=\pm 1}\equiv \frac{1}{2}\left[{\cal O}_{VV} \pm {\cal O}^F_{VV}\right]-\frac{1}{2}\left[{\cal O}_{AA} \pm {\cal O}^F_{AA}\right],\nonumber\\[0.4ex]
Q_3^{S=\pm 1}\equiv \frac{1}{2}\left[{\cal O}_{SS} \pm {\cal O}^F_{SS}\right]-\frac{1}{2}\left[{\cal O}_{PP} \pm {\cal O}^F_{PP}\right],\nonumber\\[0.4ex]
Q_4^{S=\pm 1}\equiv \frac{1}{2}\left[{\cal O}_{SS} \pm {\cal O}^F_{SS}\right]+\frac{1}{2}\left[{\cal O}_{PP} \pm {\cal O}^F_{PP}\right],\nonumber\\[0.4ex]
Q_5^{S=\pm 1}\equiv \frac{1}{2}\left[{\cal O}_{TT} \pm {\cal O}^F_{TT}\right],\nonumber
\end{cases}
\eea
\end{minipage}
\hfill
\begin{minipage}{8cm}
\bea
&\begin{cases}
{\cal Q}_1^{S=\pm 1}\equiv \frac{1}{2}\left[{\cal O}_{VA} \pm {\cal O}^F_{VA}\right]+\frac{1}{2}\left[{\cal O}_{AV} \pm {\cal O}^F_{AV}\right],\nonumber
\end{cases}\\
&\begin{cases}
{\cal Q}_2^{S=\pm 1}\equiv \frac{1}{2}\left[{\cal O}_{VA} \pm {\cal O}^F_{VA}\right]-\frac{1}{2}\left[{\cal O}_{AV} \pm {\cal O}^F_{AV}\right],\nonumber\\
{\cal Q}_3^{S=\pm 1}\equiv \frac{1}{2}\left[{\cal O}_{PS} \pm {\cal O}^F_{PS}\right]-\frac{1}{2}\left[{\cal O}_{SP} \pm {\cal O}^F_{SP}\right],\nonumber
\end{cases}\\
&\begin{cases}
{\cal Q}_4^{S=\pm 1}\equiv \frac{1}{2}\left[{\cal O}_{PS} \pm {\cal O}^F_{PS}\right]+\frac{1}{2}\left[{\cal O}_{SP} \pm {\cal O}^F_{SP}\right],\nonumber\\
{\cal Q}_5^{S=\pm 1}\equiv \frac{1}{2}\left[{\cal O}_{T\tilde T} \pm {\cal O}^F_{T\tilde T}\right].
\end{cases}\\
\label{Q_definitions}
\eea
\end{minipage}
}} 

\phantom{-------}
\noindent Summation over all independent Lorentz indices (if any),
of the Dirac matrices, is implied. The operators shown above are
grouped together according to their mixing pattern. This implies that
the renormalization matrices $Z^{S=\pm 1}$ (${\cal Z}^{S=\pm 1}$), for
the Parity Conserving (Violating) operators, have the form:
{\small{
\be
Z^{S=\pm 1}
=
\left(\begin{array}{rrrrr}
Z_{11}\,\, & Z_{12}\,\, & Z_{13}\,\, & Z_{14}\,\, & Z_{15} \\
Z_{21}\,\, & Z_{22}\,\, & Z_{23}\,\, & Z_{24}\,\, & Z_{25} \\
Z_{31}\,\, & Z_{32}\,\, & Z_{33}\,\, & Z_{34}\,\, & Z_{35} \\
Z_{41}\,\, & Z_{42}\,\, & Z_{43}\,\, & Z_{44}\,\, & Z_{45} \\
Z_{51}\,\, & Z_{52}\,\, & Z_{53}\,\, & Z_{54}\,\, & Z_{55} 
\end{array}\right)^{S=\pm 1},
\quad
{\cal Z}^{S=\pm 1}
=
\left(\begin{array}{rrrrr}
 {\cal Z}_{11}  &0\,\,         &0\,\,         &0\,\,        &0\,\,  \\
 0\,\,         &{\cal Z}_{22}  &{\cal Z}_{23}  &0\,\,        &0\,\,  \\
 0\,\,         &{\cal Z}_{32}  &{\cal Z}_{33}  &0\,\,        &0\,\,  \\
 0\,\,         &0\,\,         &0\,\,         &{\cal Z}_{44}  &{\cal Z}_{45} \\
 0\,\,         &0\,\,         &0\,\,         &{\cal Z}_{54}  &{\cal Z}_{55}
\end{array}\right)^{S=\pm 1}.
\ee
}}

Now the renormalized Parity Conserving (Violating) operators,
$\hat{Q}^{S=\pm 1}$ ($\hat{\cal Q}^{S=\pm 1}$), are defined via the
equations:
\be
{\hat Q}_l^{S=\pm 1} = Z^{S=\pm 1}_{lm} \cdot Q^{S=\pm 1}_{m} ,\quad
\hat{\cal Q}^{S=\pm 1}_l = {\cal Z}^{S=\pm 1}_{lm} \cdot {\cal Q}^{S=\pm 1}_m,
\ee
where $l,m = 1,\dots ,5$ (a sum over $m$ is implied).
The renormalized amputated Green's functions $\hat{L}^{S=\pm 1}$
($\hat{\cal L}^{S=\pm 1}$) corresponding to $Q^{S=\pm 1}$ (${\cal Q}^{S=\pm 1}$), are given
in terms of their bare counterparts $L^{S=\pm 1}$ (${\cal L}^{S=\pm 1}$) through:
\be
{\hat L}_l^{S=\pm 1} = Z_\Psi^{-2} Z^{S=\pm 1}_{lm} \cdot L^{S=\pm 1}_{m},\quad
\hat{\cal L}^{S=\pm 1}_l = Z_\Psi^{-2} {\cal Z}^{S=\pm 1}_{lm} \cdot {\cal L}^{S=\pm 1}_{m},
\label{rengreen}
\ee
where $Z_\Psi$ is the quark field renormalization constant.
In order to obtain $Z_{\Psi}$ for a given renormalization
prescription, one must make use of the inverse fermion propagator,
$S^{-1}$, calculated (up to 1-loop and up to ${\cal O}(a^2)$ for
massless Wilson/clover fermions and Symanzik improved gluons) in
Ref. \cite{CLPS}.

The renormalization matrices $Z^{S=\pm 1}$ (${\cal Z}^{S=\pm 1}$), are
computed using the appropriate Parity Conserving (Violating)
Projectors $P^{S=\pm 1}$ (${\cal P}^{S=\pm 1}$):

\begin{minipage}{6cm}
\bea
P_1^{S=\pm 1}&\equiv& + \frac{\Pi_{VV}+\Pi_{AA}}{64N_c(N_c\pm 1)},\nonumber\\
P_2^{S=\pm 1}&\equiv& + \frac{\Pi_{VV}-\Pi_{AA}}{64(N_c^2-1)} 
               \pm \frac{\Pi_{SS}-\Pi_{PP}}{32N_c(N_c^2-1)} ,\nonumber\\
P_3^{S=\pm 1}&\equiv&  \pm \frac{\Pi_{VV}-\Pi_{AA}}{32N_c(N_c^2-1)}
               +  \frac{\Pi_{SS}-\Pi_{PP}}{16(N_c^2-1)},\nonumber\\
P_4^{S=\pm 1}&\equiv& + \frac{\Pi_{SS}+\Pi_{PP}}{\frac{32N_c(N_c^2-1)}{2N_c\pm 1}}
                \mp \frac{\Pi_{TT}}{32N_c(N_c^2-1)},\nonumber\\
P_5^{S=\pm 1}&\equiv& \mp \frac{\Pi_{SS}+\Pi_{PP}}{32N_c(N_c^2-1)}
               + \ \frac{\Pi_{TT}}{\frac{96N_c(N_c^2-1)}{2N_c\mp 1}} ,\nonumber
\eea
\end{minipage}
\hfill
\begin{minipage}{7.5cm}
\bea
{\cal P}_1^{S=\pm 1}&\equiv& - \frac{\Pi_{VA}+\Pi_{AV}}{64N_c(N_c\pm 1)},\nonumber\\
{\cal P}_2^{S=\pm 1}&\equiv& - \frac{\Pi_{VA}-\Pi_{AV}}{64(N_c^2-1)}
                 \mp \frac{\Pi_{SP}-\Pi_{PS}}{32N_c(N_c^2-1)},\nonumber\\
{\cal P}_3^{S=\pm 1}&\equiv& \mp \frac{\Pi_{VA}-\Pi_{AV}}{32N_c(N_c^2-1)}
                    -  \frac{\Pi_{SP}-\Pi_{PS}}{16(N_c^2-1)},\nonumber\\
{\cal P}_4^{S=\pm 1}&\equiv& + \frac{\Pi_{SP}+\Pi_{PS}}{\frac{32N_c(N_c^2-1)}{2N_c\pm 1}}
                 \mp \frac{\Pi_{T\tilde T}}{32N_c(N_c^2-1)} ,\nonumber\\
{\cal P}_5^{S=\pm 1}&\equiv& \mp \frac{\Pi_{SP}+\Pi_{PS}}{32N_c(N_c^2-1)}
                   + \frac{\Pi_{T\tilde T}}{\frac{96N_c(N_c^2-1)}{2N_c\mp 1}} ,\nonumber
\eea
\end{minipage}
%

\phantom{-------}
\noindent where $\Pi_{XY}\equiv (X_{i_2 i_1} \otimes Y_{i_4 i_3})\delta_{a_2 a_1}\delta_{a_4 a_3}$. 
Again, summation is implied over all independent Lorentz indices (if any) of the
Dirac matrices. The above Projectors are chosen to obey
the following orthogonality conditions:
\be
{\rm Tr}(P_l^{S=\pm 1} \cdot L_{m\,(tree)}^{S=\pm 1})=\delta_{l m} ,\quad
{\rm Tr}({\cal P}_l^{S=\pm 1} \cdot {\cal L}_{m\,(tree)}^{S=\pm 1})=\delta_{l m}, 
\ee
where the trace is taken over spin and color indices, and 
${L}_{(tree)}^{S=\pm 1}$, ${\cal L}_{(tree)}^{S=\pm 1}$
are the tree-level amputated Green's functions of the operators 
${Q}^{S=\pm 1}$, ${\cal Q}^{S=\pm 1}$ respectively.

Consistently with the RI$'$ schemes, one may impose the renormalization conditions:
\be
{\rm Tr}(P_l^{S=\pm 1} \cdot {\hat L}_{m}^{S=\pm 1})=\delta_{l m} ,\quad
{\rm Tr}({\cal P}_l^{S=\pm 1} \cdot \hat{\cal L}_{m}^{S=\pm 1})=\delta_{l m}.
\label{rencon} 
\ee

These conditions should be imposed at a given renormalization scale,
$\mu$. Note, however, that due to the presence of Lorentz
non-invariant quantities, such as $\sum_\rho p_\rho^4$, which enter
the Green’s functions at ${\cal O}(a^2)$ and beyond, the
renormalization factors computed in the RI$'$-MOM scheme are also
affected through finite cutoff effects by the choice of the direction
for the external momentum.

By inserting  Eqs. (\ref{rengreen}) in the above relations, we obtain
the renormalization matrices $Z^{S=\pm 1}$, ${\cal Z}^{S=\pm 1}$ in
terms of known quantities:
\be
Z^{S=\pm 1} = Z_{\Psi}^2 \left[ \left( D^{S=\pm 1} \right)^T \right]^{-1},\quad
{\cal Z}^{S=\pm 1} = Z_{\Psi}^2 \left[ \left( {\cal D}^{S=\pm 1}
  \right)^T \right]^{-1},
\label{Zfactors}
\ee 
where:
\be
D^{S=\pm 1}_{l m} \equiv {\rm Tr}(P_l^{S=\pm 1} \cdot {L}_{m}^{S=\pm 1}) ,\quad
{\cal D}^{S=\pm 1}_{l m} \equiv {\rm Tr}({\cal P}_l^{S=\pm 1} \cdot {\cal L}_{m}^{S=\pm 1}) . 
\label{PCPV}
\ee

\noindent Note that $D^{S=\pm 1}$ and ${\cal D}^{S=\pm 1}$ have the same matrix
structure as $Z^{S=\pm 1}$ and ${\cal Z}^{S=\pm 1}$ respectively.
For convenience we express them as:

\be
D^{S=\pm1}
=\openone + \frac{g^2}{16\,\pi^2}\,
\left(\begin{array}{rrrrr}
d^\pm_{11} \hspace{0.25cm}  &d^\pm_{12}\hspace{0.25cm}   &d^\pm_{13}\hspace{0.25cm}  &d^\pm_{14}\hspace{0.25cm}   &d^\pm_{15}\hspace{0.25cm}  \\
d^\pm_{21} \hspace{0.25cm}  &d^\pm_{22}\hspace{0.25cm}   &d^\pm_{23}\hspace{0.25cm}  &d^\pm_{24}\hspace{0.25cm}   &d^\pm_{25}\hspace{0.25cm}  \\
d^\pm_{31} \hspace{0.25cm}  &d^\pm_{32}\hspace{0.25cm}   &d^\pm_{33}\hspace{0.25cm}  &d^\pm_{34}\hspace{0.25cm}   &d^\pm_{35}\hspace{0.25cm}  \\
d^\pm_{41} \hspace{0.25cm}  &d^\pm_{42}\hspace{0.25cm}   &d^\pm_{43}\hspace{0.25cm}  &d^\pm_{44}\hspace{0.25cm}   &d^\pm_{45}\hspace{0.25cm} \\
d^\pm_{51} \hspace{0.25cm}  &d^\pm_{52}\hspace{0.25cm}   &d^\pm_{53}\hspace{0.25cm}  &d^\pm_{54}\hspace{0.25cm}   &d^\pm_{55}\hspace{0.25cm}
\end{array}\right) + {\cal O}(g^4)
\label{D}
\ee

\be
{\cal D}^{S=\pm1}
=\openone + \frac{g^2}{16\,\pi^2}\,
\left(\begin{array}{rrrrr}
\delta^\pm_{11}\hspace{0.25cm} &0\hspace{0.55cm}                 &0\hspace{0.55cm}               &0\hspace{0.55cm}                &0\hspace{0.55cm}  \\
0\hspace{0.55cm}                &\delta^\pm_{22}\hspace{0.25cm}  &\delta^\pm_{23}\hspace{0.25cm} &0\hspace{0.55cm}                &0\hspace{0.55cm}  \\
0\hspace{0.55cm}                &\delta^\pm_{32}\hspace{0.25cm}  &\delta^\pm_{33}\hspace{0.25cm} &0\hspace{0.55cm}                &0\hspace{0.55cm}  \\
0\hspace{0.55cm}                &0\hspace{0.55cm}                 &0\hspace{0.55cm}               &\delta^\pm_{44}\hspace{0.25cm}  &\delta^\pm_{45}\hspace{0.25cm} \\
0\hspace{0.55cm}                &0\hspace{0.55cm}                 &0\hspace{0.55cm}               &\delta^\pm_{54}\hspace{0.25cm}  &\delta^\pm_{55}\hspace{0.25cm}
\end{array}\right) + {\cal O}(g^4)
\label{Dcal}
\ee

In the parity violating case, as explained in Ref.~\cite{DGMTV}
(Section 5.3), an equality holds between two pairs of matrix elements:
\bea
\delta^+_{22} &=& +\delta^-_{22}, \\
\delta^+_{23} &=& -\delta^-_{23} , \\
\delta^+_{32} &=& -\delta^-_{32}, \\
\delta^+_{33} &=& +\delta^-_{33}.
\eea

In addition, for the parity conserving projection the matrix elements
$d^+_{53}$, $d^-_{53}$ give zero at the 1-loop of perturbative
theory:
\bea
d^+_{53} &=& 0 , \\
d^-_{53} &=& 0 .
\eea

The matrix elements of Eqs.~(\ref{D})-(\ref{Dcal}) have the following
simple and generic form: 
\bea   
d^\pm_{l,m} &=&  d_{l,m}^{\pm(0,1)} + \csw\, d_{l,m}^{\pm(0,2)} + \csw^2\, d_{l,m}^{\pm(0,3)} + \lambda \, d_{l,m}^{\pm(0,4)}
+ \big( d_{l,m}^{\pm(0,5)} + \lambda\, d_{l,m}^{\pm(0,6)} \big)\ln(\Red{a^2}p^2)\nonumber\\[1.5ex]
&+&\Red{a^2}\,\Big[
p^2\big( d_{l,m}^{\pm(2,3)} + \csw\, d_{l,m}^{\pm(2,4)} + \csw^2\, d_{l,m}^{\pm(2,5)} + \lambda \, d_{l,m}^{\pm(2,6)} \big)\\[1.5ex]
&&\phantom{a^2} +p^2\ln(\Red{a^2}p^2)\big( d_{l,m}^{\pm(2,7)} + \csw\, d_{l,m}^{\pm(2,8)} + \csw^2\, d_{l,m}^{\pm(2,9)} + \lambda\, d_{l,m}^{\pm(2,10)}\big)\nonumber\\[1.5ex]
&&\phantom{a^2}+\frac{\sum_\mu p_\mu^4}{p^2}\big( d_{l,m}^{\pm(2,1)} + \lambda\, d_{l,m}^{\pm(2,2)} \big)
\Big] + {\cal O}(\Red{a^3}),\nonumber
\label{edlm}
\eea 
\bea   
\delta^\pm_{l,m} &=&  \delta_{l,m}^{\pm(0,1)} + \csw\, \delta_{l,m}^{\pm(0,2)} + \csw^2\, \delta_{l,m}^{\pm(0,3)} + \lambda \, \delta_{l,m}^{\pm(0,4)}
+ \big( \delta_{l,m}^{\pm(0,5)} + \lambda\, \delta_{l,m}^{\pm(0,6)} \big)\ln(\Red{a^2}p^2)\nonumber\\[1.5ex]
&+&\Red{a^2}\,\Big[
p^2\big( \delta_{l,m}^{\pm(2,3)} + \csw\, \delta_{l,m}^{\pm(2,4)} + \csw^2\, \delta_{l,m}^{\pm(2,5)} + \lambda \, \delta_{l,m}^{\pm(2,6)} \big)\\[1.5ex]
&&\phantom{a^2} +p^2\ln(\Red{a^2}p^2)\big( \delta_{l,m}^{\pm(2,7)} + \csw\, \delta_{l,m}^{\pm(2,8)} + \csw^2\, \delta_{l,m}^{\pm(2,9)} + \lambda\, \delta_{l,m}^{\pm(2,10)} \big)\nonumber\\[1.5ex]
&&\phantom{a^2} +\frac{\sum_\mu p_\mu^4}{p^2}\big( \delta_{l,m}^{\pm(2,1)} + \lambda\, \delta_{l,m}^{\pm(2,2)} \big)
\Big] + {\cal O}(\Red{a^3}).\nonumber
\label{edeltalm}
\eea 
The quantities $ d_{l,m}^{\pm(i,j)}$ and $\delta_{l,m}^{\pm(i,j)}$
appearing above are numerical coefficients depending on the number of
colors $N_c$ and the Symanzik parameters for each gluon action we have
considered; the index $i$ denotes the power of the lattice spacing $a$
that they multiply. Due to extremely lengthy results we provide the
quantities $ d_{l,m}^{\pm(i,j)}$ , $ \delta_{l,m}^{\pm(i,j)}$ 
(Tables~\ref{4f1} - \ref{4f8})
only for the special choices: $N_c=3$, $r_s=r_d=r_{s'}=r_{d'}=1$,  
and tree-level Symanzik improved action.
In all Tables the systematic errors in parentheses come from the
extrapolation ($L \to \infty$) over finite lattice sizes.
The full set of results is provided in the distribution package of
this paper as an ASCII file named: 4-fermi.m; the file is best perused
as Mathematica input; for notation see Appendix B.

The perturbative renormalization constants (Eqs.~(\ref{Zfactors})) can
be computed directly using the ${\cal O}(a^0)$ coefficients:
$d_{l,m}^{\pm(0,j)}$ and $\delta_{l,m}^{\pm(0,j)}$ (first line of
Eqs.~(\ref{edlm}) - (\ref{edeltalm})). The renormalization factor of
the fermion field, $Z_q$, is also required and in the RI$'$-MOM scheme
it reads~\cite{ACKPS}

\be
Z_q = 1 + \frac{g^2\,C_F}{16\,\pi^2}\Big[ \epsilon^{(1)}+ \csw\,\epsilon^{(2)} +
  \csw^2\,\epsilon^{(3)}  +  \lambda\,\epsilon^{(4)} - \lambda\,\ln(\Red{a^2}p^2)  \Big]\,,
\label{Zq}
\ee
\vspace{0.25cm}

\noindent where for the tree-level Symanzik improved action
\bea
\epsilon^{(1)}&=&-13.0232725(2)\,,\\
\epsilon^{(2)}&=&1.242202721(2)\,,\\
\epsilon^{(3)}&=&2.01542508(3)\,,\\
\epsilon^{(4)}&=&4.79200964(9)\,.
\eea 

\noindent As an example, we provide the exact expression for ${\cal
      Z}^{S=\pm 1}_{VA + AV}$, up to 1-loop approximation:

\bea
{\cal Z}^{S=+1}_{VA+AV} = 1 &-& \frac{g^2}{16\,\pi^2}\Bigg[
\delta_{1,1}^{+(0,1)} + \csw\, \delta_{1,1}^{+(0,2)} +
\csw^2\, \delta_{1,1}^{+(0,3)} + \lambda \,
\delta_{1,1}^{+(0,4)} + 2\,\ln(\Red{a^2}p^2)\Bigg] \nonumber \\[2ex]
&{+}& \frac{g^2\,C_F}{16\,\pi^2}\,2\,\Bigg[\epsilon^{(1)}+ \csw\,\epsilon^{(2)} +
  \csw^2\,\epsilon^{(3)}  +  \lambda\,\epsilon^{(4)} -
  \lambda\,\ln(\Red{a^2}p^2)\Bigg]\,.
\label{Z_VApAV}
\eea
\vspace{0.25cm}

\noindent This costant is the one relevant for the renormalization of $B_K$ in
the twisted mass/Osterwalder-Seiler approach~\cite{FR}, implemented in
Ref.~\cite{ETMC_BK}\footnote{In Ref.~\cite{ETMC_BK}, indeed, the
  lattice regularization chosen for the valence quarks $s$, $s'$, $d$,
  $d'$ is maximally twisted Wilson fermions with $r_s = r_{s'} = r_d
  =  -r_{d'}$. In this case, the parity conserving operator
  $Q_1=O_{VV+AA}$, besides being free from wrong chirality mixings,
  admits the same renormalization constant (here called ${\cal
    Z}_{VA+AV}$ ) as the operator ${\cal Q}_1=O_{VA+AV}$ regularized
  with untwisted Wilson quarks (i.e. with $r_s = r_{s'} = r_d  =
  r_{d'}+1$).}. For the tree-level Symanzik improved action adopted
in the calculation of Ref.~\cite{FR}, with $\csw=0$ and in the Landau
gauge, Eq.~(\ref{Z_VApAV}) reads: 

\be
Z_{VA+AV} = 1 - \frac{g^2}{16 \pi^2} \Bigg[ 2 \ln(\Red{a^2}p^2) + 42.3359 \Bigg]\,.
\ee
\vspace{0.25cm}

While the RI$'$-MOM scheme allows for a non-perturbative
renormalization procedure of the lattice operators, the Wilson
coefficients entering the effective weak Hamiltonian for neutral meson
mixing, both in the Standard Model and beyond, are often computed in
the $\rm{\overline{MS}}$ scheme. For convenience, we then also provide
here the formulae relating the operators renormalized in the RI$'$-MOM
to those renormalized in the $\rm{\overline{MS}}$ scheme, at the
next-to-leading order (i.e. 1-loop). This relation does not depend on
the chosen regularization and it may be conveniently computed using,
for instance, continuum dimensional regularization.

We restrict our attention to the $\Delta F=2$ Parity Conserving
operators, which are relevant for neutral meson mixing. These are the
5 operators $Q_{1,\ldots,5}^{S=+1}$ of Eq.(\ref{Q_definitions})  with
$S=+1$.\footnote{Note that in the $\Delta F=2$ case the operators with
  $S=-1$ vanish identically, since $O_{XY}^F=O_{XY}$. Moreover, there
  are only three Parity Violating operators, since ${\cal Q}_2^{S=+1}$
  and ${\cal Q}_3^{S=+1}$ also vanish.} For these operators, the
conversion from the RI$'$-MOM to the $\rm{\overline{MS}}$ scheme can be
written in the form

\begin{equation}
\left(Q^{S=+1}\right)^{\rm{\overline{MS}}}_l = \left( 1 + \frac{g^2}{16 \pi^2} \Delta r \right)_{lm} \left
(Q^{S=+1}\right)^{\rm RI'-MOM}_m
\label{eq:ritoms}
\end{equation}
\vspace{0.25cm}

\noindent where $\Delta r$ is a $5 \times 5$ matrix. This matrix is independent
of the choice of the regularization, i.e. it is the same for instance
for continuum dimensional regularization and for the lattice
regularization. The chiral symmetry of QCD also implies that the same
matrix $\Delta r$ is also valid for the Parity Violating sector
(though the operators ${\cal Q}_2$ and ${\cal Q}_2$ vanish in the
$\Delta F=2$ case).

When dealing with four-fermion operators, the (modified) minimal subtraction prescription in 
dimensional regularization is not sufficient however to univocally specify the renormalization 
scheme. Different $\rm{\overline{MS}}$ schemes can be defined, which differ for the definition of 
the so called evanescent operators. The scheme usually adopted in the analysis of $K-\bar K$ 
mixing is the $\rm{\overline{MS}}$ scheme defined for instance in
Ref. \cite{Buras:2000if}, for 
which the 1-loop conversion matrix $\Delta r$ of Eq. (\ref{eq:ritoms}) reads:

\begin{equation}
\Delta r = \left(
\renewcommand{\arraystretch}{1.35}
\begin{array}{ccccc}
-\frac{14}{3} + 8 \ln 2 & 0 & 0 & 0 & 0  \\
0 & -\frac{2}{3} - \frac{2}{3} \ln 2 & -4 - 4 \ln 2 & 0 &  0  \\
0 & 1 - \ln 2 & \frac{34}{3} - \frac{2}{3} \ln 2 & 0 & 0  \\
0 & 0 & 0 & \frac{10}{3} + \frac{10}{3} \ln 2 & -\frac{1}{18} + \frac{7}{18} \ln 2  \\ 
0 & 0 & 0 & \frac{56}{3} \ln 2 & -\frac{16}{9} + \frac{58}{9} \ln 2
\end{array}
\renewcommand{\arraystretch}{1.0}
\right) \ .
\label{eq:drburas}
\end{equation}
\vspace{0.25cm}

For $B-\bar B$ mixing, instead, the $\rm{\overline{MS}}$ scheme of
Ref. \cite{Beneke:1998sy} is 
more commonly adopted. The corresponding matrix $\Delta r$ differ from the one given in Eq. 
(\ref{eq:drburas}) only in the $Q_{4,5}^{S=+1}$ sector, which in this case reads

\begin{equation}
\left(\Delta r\right)_{Q_4-Q_§5} = \left(
\renewcommand{\arraystretch}{1.4}
\begin{array}{cc}
\frac{43}{6} + \frac{10}{3} \ln 2 & -\frac{7}{72} + \frac{7}{18} \ln 2  \\ 
\frac{58}{3} + \frac{56}{3} \ln 2 & -\frac{65}{18} + \frac{58}{9} \ln 2
\end{array}
\renewcommand{\arraystretch}{1.0}
\right) \ .
\label{eq:drbeneke}
\end{equation}
\vspace{0.25cm}

In order to correct, to ${\cal O}(a^2)$, non-perturbative estimates for the
renormalization constants of 4-fermion operators one should take into
account the ${\cal O}(a^2)$ corrections of Eqs.~(\ref{edlm}) -
(\ref{edeltalm}), as well as the ${\cal O}(a^2)$ terms of the fermion
propagator~\cite{CLPS}. The exact terms that need to be subtracted
from the non-perturbative $Z_q$, computed in the RI$'$-MOM scheme are
provided in Ref.~\cite{ACKPS} for general action parameters and in
Ref.~\cite{ETMC} for tree-level Symanzik improved gluons, $\csw=0$, Landau gauge.

\section{Conclusion}

The calculations presented regard all 4-fermion operators of the form:
$\bar{s}\,X\,d\,\bar{s}'\,Y\,d'$ where $X,\,Y$ are generic Dirac
matrices. Our results have explicit dependence on: $p,\,a,\,g,\,\csw,\,\lambda,\,N_c,$ and implicit
depencence on the Symanzik parameters, $c_i$. Thus, the numerical
results are presented for a selection of currently used values of
$c_i$. 

In a recent paper~\cite{ETMC_BK} the present results on the perturbative
renormalization of $\Delta F=2$ operators have been combined with
numerical simulation data in order to determine non-perturbative
renormalization coefficients with better precision. This allowed us
to extract physical values for $B_K$ with reduced lattice
artifacts.

\vskip 0.5cm
\centerline{------------------------------------------------------}
{\bf{Acknowledgements:}}
Work supported in part by the
    Research Promotion Foundation of Cyprus (Proposal Nr:
    TEXN/0308/17, $\rm ENI\Sigma X$/0506/17).

\vskip 0.5cm
\appendix
\section{Analytic expressions}
\label{appA}

In general, the final 1-loop expressions for $\Lambda_{d_1}^{XY}$, $\Lambda_{d_2}^{XY}$
and $\Lambda_{d_3}^{XY}$, up to ${\cal O}(a^2)$, are obtained as a
function of: the coupling constant $g$, clover
parameter $c_{SW}$, number of colors $N_c$, lattice spacing $a$, external
momentum $p$, and gauge parameter $\lambda$. The specific values
$\lambda=1\,(0)$ correspond to the Feynman (Landau) gauge.
Here we present the results for
$\Lambda_{d_1}^{XY}$, $\Lambda_{d_2}^{XY}$ and $\Lambda_{d_3}^{XY}$
for the special choices: 
$r_s=r_d=r_{s'}=r_{d'}=1$, and tree-level Symanzik improved gluon action.

%
\newpage
\begin{center}
\underline{\Red{Diagram d1}}
\end{center}

{\footnotesize{
\be
\hspace{-.3cm}\Blue{\Lambda_{d_1}^{XY}(p)_{i_1\,i_2\,i_3\,i_4}^{a_1\,a_2\,a_3\,a_4}}=\frac{g^2}{16\pi^2}
\left(\delta_{a_1\,a_4}\delta_{a_3\,a_2}-\frac{\delta_{a_1\,a_2}\delta_{a_3\,a_4}}{N_c}\right)
\times
\bigg \{\Blue{(\Lambda_{{\cal O}(a^0)})^{XY}_{d_1}}
+\Red{a^1}\Blue{(\Lambda_{{\cal O}(a^1)})^{XY}_{d_1}}
+\Red{a^2}\Blue{(\Lambda_{{\cal O}(a^2)})^{XY}_{d_1}}
\bigg \} ,
\ee
}}

{\footnotesize{
\bea
&&\hspace{-2.4cm}\Blue{(\Lambda_{{\cal O}(a^0)})^{XY}_{d_1}}= 
X_{i_1\,i_2}Y_{i_3\,i_4}\Big[-0.05294144(3) + 0.737558970(1)\csw + 0.238486988(3)\csw^2 \nonumber \\[1.5ex]
&&\hspace{-2.4cm}\hspace{3.8cm}
-2.100573331(5)\,\lambda +\frac{1}{2}(-1+\lambda)\ln(\Red{a^2}p^2)\Big]\nonumber \\[1.5ex]
&&\hspace{-2.4cm}+\sum_\mu (X\gamma^\mu)_{i_1\,i_2}(Y\gamma^\mu)_{i_3\,i_4}\Big[-0.507914049(6) + 0.55316919(1)\csw -0.194516637(3)\csw^2\Big]\nonumber\\[1.5ex]
&&\hspace{-2.4cm}+ \sum_{\mu,\nu} (X\gamma^\mu\gamma^\nu)_{i_1\,i_2}(Y\gamma^\mu\gamma^\nu)_{i_3\,i_4}\Big[0.018598520(2) -0.1843897425(8)\csw -0.0596217473(8)\csw^2\nonumber\\[1.5ex]
&&\hspace{-2.4cm}\hspace{4.6cm}
+\frac{1}{8}\ln(\Red{a^2}p^2)\Big] \nonumber\\[1.5ex]
&&\hspace{-2.4cm}+ \sum_{\mu,\nu,\rho} (X\gamma^\mu\gamma^\rho)_{i_1\,i_2}(Y\gamma^\nu\gamma^\rho)_{i_3\,i_4}\Big[
\frac{p_\mu p_\nu}{p^2}\Big(0.397715726853 + 0.147715726853\,\lambda \Big )\Big],
\eea
}}

{\footnotesize{
\bea
&&\hspace{-0.9cm} \Blue{(\Lambda_{{\cal O}(a^1)})^{XY}_{d_1}} = 
\sum_\mu\Big( X_{i_1\,i_2}(Y\gamma^\mu)_{i_3\,i_4} + (X\gamma^\mu)_{i_1\,i_2} \,Y_{i_3\,i_4}\Big)\times\Big[ip_\mu\Big(
0.09460083(1) -0.065711182(4)\csw \nonumber \\[1.5ex]
&&\hspace{-0.9cm} \hspace{9.4cm}
-0.059929106(1)\csw^2 +0.438508366(3)\,\lambda \nonumber \\[1.5ex]
&&\hspace{-0.9cm} \hspace{9.4cm}
+\frac{1}{4}(-1 +\csw -\lambda )\ln(\Red{a^2}p^2)
\Big)\Big]\nonumber \\[1.5ex]
&&\hspace{-0.9cm} + \sum_{\mu,\nu}\Big( (X\gamma^\mu\gamma^\nu)_{i_1\,i_2} (Y\gamma^\nu)_{i_3\,i_4}+(X\gamma^\nu)_{i_1\,i_2}(Y\gamma^\mu\gamma^\nu)_{i_3\,i_4}\Big)\times\Big[ip_\mu\Big(
0.1692905881(6) +0.010283104(5)\csw \nonumber \\[1.5ex]
&&\hspace{-0.9cm}\hspace{9.6cm}
-0.0680031615(8)\csw^2 +0.073857863427\,\lambda \nonumber \\[1.5ex]
&&\hspace{-0.9cm}\hspace{9.6cm}
+\frac{1}{16}(1+3\csw)\ln(\Red{a^2}p^2)
\Big)\Big]\nonumber \\[1.5ex]
&&\hspace{-0.9cm} + \sum_{\mu,\nu,\rho}\Big( (X\gamma^\mu\gamma^\nu\gamma^\rho)_{i_1\,i_2} (Y\gamma^\nu\gamma^\rho)_{i_3\,i_4}+(X\gamma^\nu\gamma^\rho)_{i_1\,i_2}(Y\gamma^\mu\gamma^\nu\gamma^\rho)_{i_3\,i_4}\Big)\times\Big[ip_\mu\Big(
-0.0279443091(3)\csw \nonumber \\[1.5ex]
&&\hspace{-0.9cm}\hspace{11.2cm}
+ 0.0319830668(5)\csw^2 \nonumber \\[1.5ex]
&&\hspace{-0.9cm}\hspace{11.2cm}
-\frac{1}{16}\csw \ln(\Red{a^2}p^2)
\Big)\Big],
\eea
}}
{\footnotesize{
\bea
&&\hspace{-1.3cm}\Blue{(\Lambda_{{\cal O}(a^2)})^{XY}_{d_1}} = X_{i_1\,i_2}Y_{i_3\,i_4}\Big[
p^2\Big(
1.32362251(5) -0.43684285(3)\csw -0.0208665277(5)\csw^2 \nonumber \\[1.5ex]
&&\hspace{-1.3cm} \hspace{4.3cm}+0.64073441(3)\,\lambda +\frac{1}{72}(-17 +9\csw -9\,\lambda ) \ln(\Red{a^2}p^2)  
\Big)\nonumber \\[1.5ex]
&&\hspace{-1.3cm} \hspace{3.8cm}+\frac{\sum_\sigma p_\sigma^4}{p^2}\Big(
0.06213648(8)-0.07400055(8)\,\lambda 
\Big) 
\Big]\nonumber \\[1.5ex]
&&\hspace{-1.3cm} + \sum_\mu (X\gamma^\mu)_{i_1\,i_2}(Y\gamma^\mu)_{i_3\,i_4}\Big[
p^2\Big(
0.059895142(8) -0.241755150(3)\csw +0.114731816(7)\csw^2 \nonumber \\[1.5ex]
&&\hspace{-1.3cm} \hspace{4.3cm}-0.036928931713\,\lambda +\frac{1}{48}(-7 + 11\csw -4\csw^2 )\ln(\Red{a^2}p^2) 
\Big)\nonumber \\[1.5ex]
&&\hspace{-1.3cm} \hspace{3.8cm} + p_\mu^2\Big(
1.01694823(2) -0.44474062(1)\csw -0.033265121(3)\csw^2 
\Big)
\Big] \nonumber \\[1.5ex]
&&\hspace{-1.3cm} + \sum_{\mu,\nu}\Big((X\gamma^\mu\gamma^\nu)_{i_1\,i_2}Y_{i_3\,i_4}+X_{i_1\,i_2}(Y\gamma^\mu\gamma^\nu)_{i_3\,i_4}\Big)\times \Big[
\frac{p_\nu p_\mu^3}{p^2}\Big(
0.00592406(2) -0.00295805(2)\,\lambda 
\Big)
\Big] \nonumber \\[1.5ex]
&&\hspace{-1.3cm} + \sum_{\mu,\nu} (X\gamma^\mu)_{i_1\,i_2} (Y\gamma^\nu)_{i_3\,i_4} \Big[
p_\mu p_\nu \Big(
-0.19915360(1) +0.212823513(3)\csw +0.033028338(2)\csw^2 \nonumber \\[1.5ex]
&&\hspace{-1.3cm} \hspace{4.7cm} +0.1600141922(8)\,\lambda +\frac{1}{24}(-4 +5\csw +2\csw^2 -3\,\lambda) \ln(\Red{a^2}p^2) 
\Big)
\Big] \nonumber \\[1.5ex]
&&\hspace{-1.3cm} + \sum_{\mu,\nu} (X\gamma^\mu\gamma^\nu)_{i_1\,i_2} (Y\gamma^\mu\gamma^\nu)_{i_3\,i_4}\Big[
p^2\Big(
-0.08962805(1) +0.0769373498(3)\csw +0.0067184623(3)\csw^2 \nonumber \\[1.5ex]
&&\hspace{-1.3cm} \hspace{5.2cm} +\frac{1}{240} (7 -5\csw)\ln(\Red{a^2}p^2) 
\Big)\nonumber \\[1.5ex]
&&\hspace{-1.3cm} \hspace{4.6cm} +p_\mu^2\Big(
+0.16608907(6) +0.07446360(2)\csw -0.0087763322(3)\csw^2 \nonumber \\[1.5ex]
&&\hspace{-1.3cm} \hspace{5.2cm} -\frac{29}{180}\ln(\Red{a^2}p^2)
\Big)\nonumber \\[1.5ex]
&&\hspace{-1.3cm} \hspace{4.6cm} +\frac{\sum_\sigma p_\sigma^4}{p^2}\Big(
-0.048180849735
\Big)
\Big] \nonumber \\[1.5ex]
&&\hspace{-1.3cm} + \sum_{\mu,\nu,\rho} (X\gamma^\mu\gamma^\rho)_{i_1\,i_2} (Y\gamma^\nu\gamma^\rho)_{i_3\,i_4}\Big[
p_\mu p_\nu \Big(
-0.21865904(4) +0.054629909(8)\csw +0.00276900638(7)\csw^2 \nonumber \\[1.5ex]
&&\hspace{-1.3cm} \hspace{5.5cm} -0.082411837(6)\,\lambda +\frac{1}{1440}(164 -60\csw +45\,\lambda)\ln(\Red{a^2}p^2)
\Big)\nonumber \\[1.5ex]
&&\hspace{-1.3cm} \hspace{4.6cm} +\frac{(p_\mu^3 p_\nu + p_\mu p_\nu^3)}{p^2}\Big(
-0.110138789528 -0.024619287809\,\lambda
\Big)\nonumber \\[1.5ex]
&&\hspace{-1.3cm} \hspace{4.6cm} + p_\mu p_\nu \frac{\sum_\sigma p_\sigma^4}{(p^2)^2}\Big(
0.140961390102 +0.045240352404\,\lambda 
\Big)\nonumber \\[1.5ex]
&&\hspace{-1.3cm} \hspace{4.6cm} +\frac{p_\mu p_\nu p_\rho^2}{p^2}\Big(
-0.477634781(8)-0.083831642(8)\,\lambda 
\Big)
\Big] \nonumber \\[1.5ex]
&&\hspace{-1.3cm} + \sum_{\mu,\nu,\rho} (X\gamma^\mu\gamma^\nu\gamma^\rho)_{i_1\,i_2} (Y\gamma^\mu\gamma^\nu\gamma^\rho)_{i_3\,i_4} \Big[
p_\mu^2 \Big( 
0.00385492408(3)\csw^2 
\Big) 
\Big] \nonumber \\[1.5ex]
&&\hspace{-1.3cm} + \sum_{\mu,\nu,\rho,\sigma} (X\gamma^\mu\gamma^\rho\gamma^\sigma)_{i_1\,i_2} (Y\gamma^\nu\gamma^\rho\gamma^\sigma)_{i_3\,i_4} \Big[
p_\mu p_\nu \Big(
-0.0209503296(6)\csw^2 -\frac{1}{32}\csw^2 \,\ln(\Red{a^2}p^2) 
\Big) 
\Big].
\eea
}}

%
\vspace{-1.0cm}
\begin{center}
\Red{\underline{Diagram d2}}
\end{center}

{\footnotesize{
\be
\hspace{-.3cm}\Blue{\Lambda_{d_2}^{XY}(p)_{i_1\,i_2\,i_3\,i_4}^{a_1\,a_2\,a_3\,a_4}}=\frac{g^2}{16\pi^2}\delta_{a_1\,a_2}\delta_{a_3\,a_4}
\left(N_c-\frac{1}{N_c}\right)
\times
\bigg \{\Blue{(\Lambda_{{\cal O}(a^0)})^{XY}_{d_2}}+\Red{a^1}\Blue{(\Lambda_{{\cal O}(a^1)})^{XY}_{d_2}} +\Red{a^2}\Blue{(\Lambda_{{\cal O}(a^2)})^{XY}_{d_2}}
\bigg \} ,
\ee
}}

{\footnotesize{
\bea
&&\hspace{-2.4cm}\Blue{(\Lambda_{{\cal O}(a^0)})^{XY}_{d_2}}= 
X_{i_1\,i_2}Y_{i_3\,i_4}\Big[1.2904478(4) +0.737558970(1)\csw +0.238486988(3)\csw^2 \nonumber \\[1.5ex]
&&\hspace{-2.4cm}\hspace{3.8cm}
+2.3960046(4)\,\lambda +\frac{1}{2}(-1-\lambda)\ln(\Red{a^2}p^2)\Big]\nonumber \\[1.5ex]
&&\hspace{-2.4cm}+\sum_\mu X_{i_1\,i_2}(\gamma^\mu Y\gamma^\mu)_{i_3\,i_4}\Big[-0.507914047(8) + 0.55316917(2)\csw -0.194516638(9)\csw^2\Big]\nonumber\\[1.5ex]
&&\hspace{-2.4cm}+ \sum_{\mu,\nu} X_{i_1\,i_2}(\gamma^\mu\gamma^\nu Y\gamma^\mu\gamma^\nu)_{i_3\,i_4}\Big[-0.129117207(2) -0.1843897425(8)\csw -0.0596217473(8)\csw^2\nonumber\\[1.5ex]
&&\hspace{-2.4cm}\hspace{4.4cm}
+\frac{1}{8}\ln(\Red{a^2}p^2)\Big] \nonumber\\[1.5ex]
&&\hspace{-2.4cm}+ \sum_{\mu,\nu,\rho} X_{i_1\,i_2}(\gamma^\mu\gamma^\rho Y\gamma^\nu\gamma^\rho)_{i_3\,i_4}\Big[
\frac{p_\mu p_\nu}{p^2}\Big(-\frac{1}{4}\,\lambda \Big )\Big],
\eea
}}

{\footnotesize{
\bea
&&\hspace{-1.0cm} \Blue{(\Lambda_{{\cal O}(a^1)})^{XY}_{d_2}} = 
\sum_\mu\Big(X_{i_1\,i_2} \,(Y\gamma^\mu)_{i_3\,i_4} + X_{i_1\,i_2}(\gamma^\mu Y)_{i_3\,i_4}\Big)\times\Big[ip_\mu\Big(
0.37785613(9) -0.56675680(2)\csw \nonumber \\[1.5ex]
&&\hspace{-1.0cm} \hspace{9.4cm}
-0.160026205(2)\csw^2 -0.48393977(9)\,\lambda \nonumber \\[1.5ex]
&&\hspace{-1.0cm} \hspace{9.4cm}
+\frac{1}{8}(-3 +5\csw +2\,\lambda )\ln(\Red{a^2}p^2)
\Big)\Big]\nonumber \\[1.5ex]
&&\hspace{-1.0cm} + \sum_{\mu,\nu}\Big(X_{i_1\,i_2} (\gamma^\mu\gamma^\nu Y\gamma^\nu)_{i_3\,i_4}+X_{i_1\,i_2}(\gamma^\nu Y\gamma^\nu\gamma^\mu)_{i_3\,i_4}\Big)\times\Big[ip_\mu\Big(
-0.073251555(1) -0.001704761(4)\csw \nonumber \\[1.5ex]
&&\hspace{-1.0cm}\hspace{9.0cm}
+0.032093938(2)\csw^2 -\frac{1}{8}\,\lambda \nonumber \\[1.5ex]
&&\hspace{-1.0cm}\hspace{9.0cm}
+\frac{1}{16}(3-\csw)\ln(\Red{a^2}p^2)
\Big)\Big]\nonumber \\[1.5ex]
&&\hspace{-1.0cm} + \sum_{\mu,\nu,\rho}\Big( X_{i_1\,i_2} (\gamma^\mu\gamma^\nu\gamma^\rho Y\gamma^\nu\gamma^\rho)_{i_3\,i_4}+X_{i_1\,i_2}(\gamma^\nu\gamma^\rho Y\gamma^\nu\gamma^\rho\gamma^\mu)_{i_3\,i_4}\Big)\times\Big[ip_\mu\Big(
0.0459135542(5)\csw \nonumber \\[1.5ex]
&&\hspace{-1.0cm}\hspace{10.6cm}
+ 0.0319830668(5)\csw^2 \nonumber \\[1.5ex]
&&\hspace{-1.0cm}\hspace{10.6cm}
-\frac{1}{16}\csw \ln(\Red{a^2}p^2)
\Big)\Big],
\eea
}}
{\footnotesize{
\bea
&&\hspace{-1.3cm}\Blue{(\Lambda_{{\cal O}(a^2)})^{XY}_{d_2}} = X_{i_1\,i_2}Y_{i_3\,i_4}\Big[
p^2\Big(
0.7374671(6) -0.24301094(4)\csw -0.0096054476(8)\csw^2 \nonumber \\[1.5ex]
&&\hspace{-1.3cm} \hspace{4.3cm}-0.4696085(6)\,\lambda +\frac{1}{120}(-23 +5\csw +15\,\lambda ) \ln(\Red{a^2}p^2)  
\Big)\nonumber \\[1.5ex]
&&\hspace{-1.3cm} \hspace{3.8cm}+\frac{\sum_\sigma p_\sigma^4}{p^2}\Big(
\frac{1}{90}(-77 +15\,\lambda )
\Big) 
\Big]\nonumber \\[1.5ex]
&&\hspace{-1.3cm} + \sum_\mu X_{i_1\,i_2}(\gamma^\mu Y\gamma^\mu)_{i_3\,i_4}\Big[
p^2\Big(
0.04610701(4) -0.19136171(3)\csw +0.02347831(4)\csw^2 \nonumber \\[1.5ex]
&&\hspace{-1.3cm} \hspace{4.2cm}-0.06249999(1)\,\lambda +\frac{1}{16}(1 + \csw)\ln(\Red{a^2}p^2) 
\Big)\nonumber \\[1.5ex]
&&\hspace{-1.3cm} \hspace{3.7cm} + p_\mu^2\Big(
0.17251518(3) -0.19211806(3)\csw +0.01744902(2)\csw^2 
\Big)
\Big] \nonumber \\[1.5ex]
&&\hspace{-1.3cm} + \sum_{\mu,\nu}\Big(X_{i_1\,i_2}(Y\gamma^\mu\gamma^\nu)_{i_3\,i_4}+X_{i_1\,i_2}(\gamma^\nu\gamma^\mu Y)_{i_3\,i_4}\Big)\times \Big[
\frac{p_\nu p_\mu^3}{p^2}\Big(
\frac{101}{288}
\Big)
\Big] \nonumber \\[1.5ex]
&&\hspace{-1.3cm} + \sum_{\mu,\nu} X_{i_1\,i_2} (\gamma^\mu Y\gamma^\nu)_{i_3\,i_4} \Big[
p_\mu p_\nu \Big(
0.05513763(3) -0.005630284(8)\csw -0.072690409(7)\csw^2 \nonumber \\[1.5ex]
&&\hspace{-1.3cm} \hspace{4.6cm} -0.10887203(3)\,\lambda +\frac{1}{8}(-2 +\csw +\csw^2 +\lambda) \ln(\Red{a^2}p^2) 
\Big)
\Big] \nonumber \\[1.5ex]
&&\hspace{-1.3cm} + \sum_{\mu,\nu} X_{i_1\,i_2} (\gamma^\mu\gamma^\nu Y\gamma^\mu\gamma^\nu)_{i_3\,i_4}\Big[
p^2\Big(
-0.05064893(1) +0.0394316274(6)\csw +0.00332360968(9)\csw^2 \nonumber \\[1.5ex]
&&\hspace{-1.3cm} \hspace{5.0cm} +\frac{1}{720} (13 -15\csw)\ln(\Red{a^2}p^2) 
\Big)\nonumber \\[1.5ex]
&&\hspace{-1.3cm} \hspace{4.4cm} +p_\mu^2\Big(
+0.05383442(9) +0.124311493(7)\csw -0.0051958638(1)\csw^2 \nonumber \\[1.5ex]
&&\hspace{-1.3cm} \hspace{5.0cm} -\frac{1}{15}\ln(\Red{a^2}p^2)
\Big)\nonumber \\[1.5ex]
&&\hspace{-1.3cm} \hspace{4.4cm} +\frac{\sum_\sigma p_\sigma^4}{p^2}\Big(
-\frac{1}{240}
\Big)
\Big] \nonumber \\[1.5ex]
&&\hspace{-1.3cm} + \sum_{\mu,\nu,\rho} X_{i_1\,i_2} (\gamma^\mu\gamma^\rho Y\gamma^\rho\gamma^\nu)_{i_3\,i_4}\Big[
p_\mu p_\nu \Big(
-0.03270359(5) -0.039026988(4)\csw +0.0015068706(2)\csw^2 \nonumber \\[1.5ex]
&&\hspace{-1.3cm} \hspace{5.5cm} -0.06926696(4)\,\lambda +\frac{1}{1440}(28 +60\csw +45\,\lambda)\ln(\Red{a^2}p^2)
\Big)\nonumber \\[1.5ex]
&&\hspace{-1.3cm} \hspace{4.5cm} +\frac{(p_\mu^3 p_\nu + p_\mu p_\nu^3)}{p^2}\Big(
\frac{1}{960}(41-40\,\lambda) 
\Big)\nonumber \\[1.5ex]
&&\hspace{-1.3cm} \hspace{4.5cm} + p_\mu p_\nu \frac{\sum_\sigma p_\sigma^4}{(p^2)^2}\Big(
\frac{1}{960}(7+25\,\lambda) 
\Big)\nonumber \\[1.5ex]
&&\hspace{-1.3cm} \hspace{4.5cm} +\frac{p_\mu p_\nu p_\rho^2}{p^2}\Big(
\frac{1}{288}(-40-9\,\lambda) 
\Big)
\Big] \nonumber \\[1.5ex]
&&\hspace{-1.3cm} + \sum_{\mu,\nu,\rho} X_{i_1\,i_2} (\gamma^\mu\gamma^\nu\gamma^\rho Y\gamma^\mu\gamma^\nu\gamma^\rho)_{i_3\,i_4} \Big[
p_\mu^2 \Big( 
0.00385492795(2)\csw^2 
\Big) 
\Big] \nonumber \\[1.5ex]
&&\hspace{-1.3cm} + \sum_{\mu,\nu,\rho,\sigma} X_{i_1\,i_2} (\gamma^\mu\gamma^\rho\gamma^\sigma Y\gamma^\nu\gamma^\rho\gamma^\sigma)_{i_3\,i_4} \Big[
p_\mu p_\nu \Big(
0.015978597(1)\csw^2 -\frac{1}{32}\csw^2 \,\ln(\Red{a^2}p^2) 
\Big) 
\Big].
\eea
}}

%
\vspace{-1.0cm}
\begin{center}
\Red{\underline{Diagram d3}}
\end{center}

{\footnotesize{
\be
\hspace{-.3cm}\Blue{\Lambda_{d_3}^{XY}(p)_{i_1\,i_2\,i_3\,i_4}^{a_1\,a_2\,a_3\,a_4}}=\frac{g^2}{16\pi^2}
\left(\delta_{a_1\,a_4}\delta_{a_3\,a_2}-\frac{\delta_{a_1\,a_2}\delta_{a_3\,a_4}}{N_c}\right)
\times
\bigg \{\Blue{(\Lambda_{{\cal O}(a^0)})^{XY}_{d_3}}+\Red{a^1}\Blue{(\Lambda_{{\cal O}(a^1)})^{XY}_{d_3}} +\Red{a^2}\Blue{(\Lambda_{{\cal O}(a^2)})^{XY}_{d_3}}
\bigg \} ,
\ee
}}

{\footnotesize{
\bea
&&\hspace{-2.4cm}\Blue{(\Lambda_{{\cal O}(a^0)})^{XY}_{d_3}}= 
X_{i_1\,i_2}Y_{i_3\,i_4}\Big[1.2904478(4) +0.737558970(1)\csw +0.238486988(3)\csw^2 \nonumber \\[1.5ex]
&&\hspace{-2.4cm}\hspace{3.8cm}
+2.3960046(4)\,\lambda +\frac{1}{2}(-1-\lambda)\ln(\Red{a^2}p^2)\Big]\nonumber \\[1.5ex]
&&\hspace{-2.4cm}+\sum_\mu (X\gamma^\mu)_{i_1\,i_2}(\gamma^\mu Y)_{i_3\,i_4}\Big[-0.507914047(8) + 0.55316917(2)\csw -0.194516638(9)\csw^2\Big]\nonumber\\[1.5ex]
&&\hspace{-2.4cm}+ \sum_{\mu,\nu} (X\gamma^\mu\gamma^\nu)_{i_1\,i_2}(\gamma^\mu\gamma^\nu Y)_{i_3\,i_4}\Big[-0.129117207(2) -0.1843897425(8)\csw -0.0596217473(8)\csw^2\nonumber\\[1.5ex]
&&\hspace{-2.4cm}\hspace{4.6cm}
+\frac{1}{8}\ln(\Red{a^2}p^2)\Big] \nonumber\\[1.5ex]
&&\hspace{-2.4cm}+ \sum_{\mu,\nu,\rho} (X\gamma^\mu\gamma^\rho)_{i_1\,i_2}(\gamma^\nu\gamma^\rho Y)_{i_3\,i_4}\Big[
\frac{p_\mu p_\nu}{p^2}\Big(-\frac{1}{4}\,\lambda \Big )\Big],
\eea
}}

{\footnotesize{
\bea
&&\hspace{-1.0cm} \Blue{(\Lambda_{{\cal O}(a^1)})^{XY}_{d_3}} = 
\sum_\mu\Big((X\gamma^\mu)_{i_1\,i_2} \,Y_{i_3\,i_4} + X_{i_1\,i_2}(\gamma^\mu Y)_{i_3\,i_4}\Big)\times\Big[ip_\mu\Big(
0.37785613(9) -0.56675680(2)\csw \nonumber \\[1.5ex]
&&\hspace{-1.0cm} \hspace{9.4cm}
-0.160026205(2)\csw^2 -0.48393977(9)\,\lambda \nonumber \\[1.5ex]
&&\hspace{-1.0cm} \hspace{9.4cm}
+\frac{1}{8}(-3 +5\csw +2\,\lambda )\ln(\Red{a^2}p^2)
\Big)\Big]\nonumber \\[1.5ex]
&&\hspace{-1.0cm} + \sum_{\mu,\nu}\Big((X\gamma^\nu )_{i_1\,i_2} (\gamma^\mu\gamma^\nu Y)_{i_3\,i_4}+(X\gamma^\nu\gamma^\mu)_{i_1\,i_2}(\gamma^\nu Y)_{i_3\,i_4}\Big)\times\Big[ip_\mu\Big(
-0.073251555(1) -0.001704761(4)\csw \nonumber \\[1.5ex]
&&\hspace{-1.0cm}\hspace{9.6cm}
+0.032093938(2)\csw^2 -\frac{1}{8}\,\lambda \nonumber \\[1.5ex]
&&\hspace{-1.0cm}\hspace{9.6cm}
+\frac{1}{16}(3-\csw)\ln(\Red{a^2}p^2)
\Big)\Big]\nonumber \\[1.5ex]
&&\hspace{-1.0cm} + \sum_{\mu,\nu,\rho}\Big( (X\gamma^\mu\gamma^\nu\gamma^\rho )_{i_1\,i_2} (\gamma^\nu\gamma^\rho Y)_{i_3\,i_4}+(X\gamma^\nu\gamma^\rho )_{i_1\,i_2}(\gamma^\nu\gamma^\rho\gamma^\mu Y)_{i_3\,i_4}\Big)\times\Big[ip_\mu\Big(
0.0459135542(5)\csw \nonumber \\[1.5ex]
&&\hspace{-1.0cm}\hspace{10.5cm}
+ 0.0319830668(5)\csw^2 \nonumber \\[1.5ex]
&&\hspace{-1.0cm}\hspace{10.5cm}
-\frac{1}{16}\csw \ln(\Red{a^2}p^2)
\Big)\Big],
\eea
}}
{\footnotesize{
\bea
&&\hspace{-1.3cm}\Blue{(\Lambda_{{\cal O}(a^2)})^{XY}_{d_3}} = X_{i_1\,i_2}Y_{i_3\,i_4}\Big[
p^2\Big(
0.7374671(6) -0.24301094(4)\csw -0.0096054476(8)\csw^2 \nonumber \\[1.5ex]
&&\hspace{-1.3cm} \hspace{4.3cm}-0.4696085(6)\,\lambda +\frac{1}{120}(-23 +5\csw +15\,\lambda ) \ln(\Red{a^2}p^2)  
\Big)\nonumber \\[1.5ex]
&&\hspace{-1.3cm} \hspace{3.8cm}+\frac{\sum_\sigma p_\sigma^4}{p^2}\Big(
\frac{1}{90}(-77 +15\,\lambda )
\Big) 
\Big]\nonumber \\[1.5ex]
&&\hspace{-1.3cm} + \sum_\mu (X\gamma^\mu)_{i_1\,i_2}(\gamma^\mu Y)_{i_3\,i_4}\Big[
p^2\Big(
0.04610701(4) -0.19136171(3)\csw +0.02347831(4)\csw^2 \nonumber \\[1.5ex]
&&\hspace{-1.3cm} \hspace{4.3cm}-0.06249999(1)\,\lambda +\frac{1}{16}(1 + \csw)\ln(\Red{a^2}p^2) 
\Big)\nonumber \\[1.5ex]
&&\hspace{-1.3cm} \hspace{3.8cm} + p_\mu^2\Big(
0.17251518(3) -0.19211806(3)\csw +0.01744902(2)\csw^2 
\Big)
\Big] \nonumber \\[1.5ex]
&&\hspace{-1.3cm} + \sum_{\mu,\nu}\Big((X\gamma^\mu\gamma^\nu)_{i_1\,i_2}Y_{i_3\,i_4}+X_{i_1\,i_2}(\gamma^\nu\gamma^\mu Y)_{i_3\,i_4}\Big)\times \Big[
\frac{p_\nu p_\mu^3}{p^2}\Big(
\frac{101}{288}
\Big)
\Big] \nonumber \\[1.5ex]
&&\hspace{-1.3cm} + \sum_{\mu,\nu} (X\gamma^\mu )_{i_1\,i_2} (\gamma^\nu Y)_{i_3\,i_4} \Big[
p_\mu p_\nu \Big(
0.05513763(3) -0.005630284(8)\csw -0.072690409(7)\csw^2 \nonumber \\[1.5ex]
&&\hspace{-1.3cm} \hspace{4.7cm} -0.10887203(3)\,\lambda +\frac{1}{8}(-2 +\csw +\csw^2 +\lambda) \ln(\Red{a^2}p^2) 
\Big)
\Big] \nonumber \\[1.5ex]
&&\hspace{-1.3cm} + \sum_{\mu,\nu} (X\gamma^\mu\gamma^\nu)_{i_1\,i_2} (\gamma^\mu\gamma^\nu Y)_{i_3\,i_4}\Big[
p^2\Big(
-0.05064893(1) +0.0394316274(6)\csw +0.00332360968(9)\csw^2 \nonumber \\[1.5ex]
&&\hspace{-1.3cm} \hspace{5.1cm} +\frac{1}{720} (13 -15\csw)\ln(\Red{a^2}p^2) 
\Big)\nonumber \\[1.5ex]
&&\hspace{-1.3cm} \hspace{4.6cm} +p_\mu^2\Big(
+0.05383442(9) +0.124311493(7)\csw -0.0051958638(1)\csw^2 \nonumber \\[1.5ex]
&&\hspace{-1.3cm} \hspace{5.2cm} -\frac{1}{15}\ln(\Red{a^2}p^2)
\Big)\nonumber \\[1.5ex]
&&\hspace{-1.3cm} \hspace{4.6cm} +\frac{\sum_\sigma p_\sigma^4}{p^2}\Big(
-\frac{1}{240}
\Big)
\Big] \nonumber \\[1.5ex]
&&\hspace{-1.3cm} + \sum_{\mu,\nu,\rho} (X\gamma^\rho\gamma^\mu)_{i_1\,i_2} (\gamma^\nu\gamma^\rho Y)_{i_3\,i_4}\Big[
p_\mu p_\nu \Big(
-0.03270359(5) -0.039026988(4)\csw +0.0015068706(2)\csw^2 \nonumber \\[1.5ex]
&&\hspace{-1.3cm} \hspace{5.5cm} -0.06926696(4)\,\lambda +\frac{1}{1440}(28 +60\csw +45\,\lambda)\ln(\Red{a^2}p^2)
\Big)\nonumber \\[1.5ex]
&&\hspace{-1.3cm} \hspace{4.7cm} +\frac{(p_\mu^3 p_\nu + p_\mu p_\nu^3)}{p^2}\Big(
\frac{1}{960}(41-40\,\lambda) 
\Big)\nonumber \\[1.5ex]
&&\hspace{-1.3cm} \hspace{4.7cm} + p_\mu p_\nu \frac{\sum_\sigma p_\sigma^4}{(p^2)^2}\Big(
\frac{1}{960}(7+25\,\lambda) 
\Big)\nonumber \\[1.5ex]
&&\hspace{-1.3cm} \hspace{4.7cm} +\frac{p_\mu p_\nu p_\rho^2}{p^2}\Big(
\frac{1}{288}(-40-9\,\lambda) 
\Big)
\Big] \nonumber \\[1.5ex]
&&\hspace{-1.3cm} + \sum_{\mu,\nu,\rho} (X\gamma^\mu\gamma^\nu\gamma^\rho)_{i_1\,i_2} (\gamma^\mu\gamma^\nu\gamma^\rho Y)_{i_3\,i_4} \Big[
p_\mu^2 \Big( 
0.00385492795(2)\csw^2 
\Big) 
\Big] \nonumber \\[1.5ex]
&&\hspace{-1.3cm} + \sum_{\mu,\nu,\rho,\sigma} (X\gamma^\mu\gamma^\rho\gamma^\sigma)_{i_1\,i_2} (\gamma^\nu\gamma^\rho\gamma^\sigma Y)_{i_3\,i_4} \Big[
p_\mu p_\nu \Big(
0.015978597(1)\csw^2 -\frac{1}{32}\csw^2 \,\ln(\Red{a^2}p^2) 
\Big) 
\Big].
\eea
}}

\section{Notation in ASCII file: 4-fermi.m }
\label{appB}

The full body of our results can be accessed online through the
file 4-fermi.m, which is a Mathematica input file. It includes
the expressions for the three Feynman diagrams:
\vspace{-0.25cm}
\begin{itemize}
\item $\Lambda_{d_1}^{XY}$: d1[action,csw,lambda,Nc,g,aL]
\item $\Lambda_{d_2}^{XY}$: d2[action,csw,lambda,Nc,g,aL]
\item $\Lambda_{d_3}^{XY}$: d3[action,csw,lambda,Nc,g,aL]
\end{itemize}
from which one can construct the matrix elements of any 4-fermion
operator of the above form. Each expression depends on the variables:
\vspace{-0.25cm}
\begin{itemize}
\item action: Selection of improved gauge action as follows, 1 $\rightarrow$
  Plaquette, 2 $\rightarrow$ Tree Level Symanzik, 3 $\rightarrow$ TILW
  ($\beta\,c_0=8.60$), 4 $\rightarrow$ TILW ($\beta\,c_0=8.45$), 5
  $\rightarrow$ TILW ($\beta\,c_0=8.30$), 6 $\rightarrow$ TILW
  ($\beta\,c_0=8.20$), 7 $\rightarrow$ TILW ($\beta\,c_0=8.10$), 8
  $\rightarrow$ TILW ($\beta\,c_0=8.00$), 9 $\rightarrow$ Iwasaki, 10
  $\rightarrow$ DBW2   
\vspace{-0.25cm}\item csw: clover parameter
\vspace{-0.25cm}\item lambda: gauge parameter (Landau/Feynman/Generic correspond to
  0/1/lambda)
\vspace{-0.25cm}\item Nc: number of colors                                            
\vspace{-0.25cm}\item g: coupling constant
\vspace{-0.25cm}\item aL: lattice spacing   
\end{itemize}
In particular, the quantities of interest in that file are the
renornalization matrices for the 10 Parity Conserving operators, and
10 Parity Violating operators, which read: 
\begin{itemize}
\vspace{-0.15cm}\item $Z^{S=+1}$: PCplus[action,csw,lambda,Nc,g,aL,p2,p4][Projector,LGreen]
\vspace{-0.15cm}\item $Z^{S=-1}$: PCminus[action,csw,lambda,Nc,g,aL,p2,p4][Projector,LGreen]
\vspace{-0.15cm}\item ${\cal Z}^{S=+1}$: PVplus[action,csw,lambda,Nc,g,aL,p2,p4][Projector,LGreen]
\vspace{-0.15cm}\item ${\cal Z}^{S=-1}$: PVminus[action,csw,lambda,Nc,g,aL,p2,p4][Projector,LGreen] .
\end{itemize}
The additinal variables are 
\vspace{-0.25cm}
\begin{itemize}
\item p2: $\sum_{i=1}^4 p_i^2$
\item p4: $\sum_{i=1}^4 p_i^4$
\item Projector: the index $l$ of Section~\ref{Sec_renorm} (1 to 5)
\vspace{-0.25cm}\item LGreen: the index $m$ of Section~\ref{Sec_renorm} (1 to 5)
\end{itemize}


\newpage
\begin{turnpage}
\renewcommand{\arraystretch}{0.65}

\begin{table}[!h]
{\small{
\begin{center}
\begin{minipage}{24.5cm}
\begin{tabular}{|l|r@{}l|r@{}l|r@{}l|r@{}l|r@{}l|r@{}l|r@{}l|r@{}l|}
\multicolumn{17}{c}{\Red{Parity Conserving, Flavour Exchange Symmetry Plus }}\\
\hline
\hline
\multicolumn{1}{|c|}{$(l,\,m)$}&
\multicolumn{2}{c|}{$ d_{l,m}^{{+(0,1)}^{\phantom{A}}}$} &
\multicolumn{2}{c|}{$ d_{l,m}^{{+(0,2)}^{\phantom{A}}}$} &
\multicolumn{2}{c|}{$ d_{l,m}^{{+(0,3)}^{\phantom{A}}}$} &
\multicolumn{2}{c|}{$ d_{l,m}^{{+(0,4)}^{\phantom{A}}}$} &
\multicolumn{2}{c|}{$ d_{l,m}^{{+(0,5)}^{\phantom{A}}}$} &
\multicolumn{2}{c|}{$ d_{l,m}^{{+(0,6)}^{\phantom{A}}}$} &
\multicolumn{2}{c|}{$ d_{l,m}^{{+(2,1)}^{\phantom{A}}}$} &
\multicolumn{2}{c|}{$ d_{l,m}^{{+(2,2)}^{\phantom{A}}}$}\\
\hline
\hline
$(1,\,1)$    &7.&607190(2)      &-2.&95023588(3)    &-0.&95394796(3)    &12.&293750(2)  &2&             &-8/&3          &-2.&79899092(5)   &0.&39295395(5)  \\  
$(1,\,2)$    &5.&41774985(8)    &-5.&9004712(3)     &2.&0748441(1)      &0&             &0&             &0&             &0&                &0&              \\ 
$(1,\,3)$    &-0.&67721873(1)   &0.&73755889(5)     &-0.&25935552(2)    &0&             &0&             &0&             &0&                &0&              \\ 
$(1,\,4)$    &-0.&67721873(1)   &0.&73755892(2)     &-0.&259355516(6)   &0&             &0&             &0&             &0&                &0&              \\ 
$(1,\,5)$    &-2.&03165620(4)   &2.&21267677(7)     &-0.&77806655(2)    &0&             &0&             &0&             &0&                &0&              \\ 
\hline                                                                                                                                                                  
$(2,\,1)$    &7.&4494060(1)     &-8.&1131478(4)     &2.&8529107(1)      &0&             &0&             &0&             &0&                &0&              \\ 
$(2,\,2)$    &4.&331175(2)      &-3.&68779471(5)    &0.&30109257(1)     &10.&816593(2)  &1&             &-8/&3          &-0.&87361423(2)   &0.&57016434(2)  \\  
$(2,\,3)$    &-0.&8545378(4)    &-0.&368779461(8)   &0.&129677758(2)    &-0.&1931470(4) &0&             &0&             &0.&89270365(2)    &0.&08962151(2)  \\ 
$(2,\,4)$    &0.&338609366(5)   &-0.&36877945(2)    &0.&129677758(6)    &0&             &0&             &0&             &0&                &0&              \\ 
$(2,\,5)$    &-1.&01582810(1)   &1.&10633834(5)     &-0.&38903328(2)    &0&             &0&             &0&             &0&                &0&              \\ 
\hline                                                                                                                                                                  
$(3,\,1)$    &1.&35443746(2)    &-1.&47511779(6)    &0.&51871103(2)     &0&             &0&             &0&             &0&                &0&              \\ 
$(3,\,2)$    &-9.&615778(1)     &-10.&32582548(4)   &-2.&34313283(3)    &-3.&772588(1)  &6&             &0&             &1.&70275904(9)    &0.&92098603(9)  \\ 
$(3,\,3)$    &13.&627614(2)     &9.&58826675(6)     &4.&59385837(5)     &15.&316593(2)  &-8&            &-8/&3          &1.&92846910(2)    &-0.&27358566(2) \\  
$(3,\,4)$    &-8.&8038435(2)    &9.&5882656(6)      &-3.&3716217(2)     &0&             &0&             &0&             &0&                &0&              \\ 
$(3,\,5)$    &-6.&09496858(8)   &6.&6380301(3)      &-2.&3341997(1)     &0&             &0&             &0&             &0&                &0&              \\ 
\hline                                                                                                                                                                  
$(4,\,1)$    &-2.&70887493(5)   &2.&9502357(1)      &-1.&03742206(3)    &0&             &0&             &0&             &0&                &0&              \\ 
$(4,\,2)$    &9.&4810622(1)     &-10.&3258245(4)    &3.&6309772(1)      &0&             &0&             &0&             &0&                &0&              \\ 
$(4,\,3)$    &-10.&8354997(2)   &11.&8009423(6)     &-4.&1496883(2)     &0&             &0&             &0&             &0&                &0&              \\ 
$(4,\,4)$    &10.&269733(2)     &7.&37558970(4)     &2.&38486989(5)     &14.&816594(2)  &-5&            &-8/&3          &2.&01567488(3)    &0.&02849782(3)  \\  
$(4,\,5)$    &9.&732709(2)      &7.&37558970(3)     &2.&38486988(5)     &3.&477157(2)   &-5&            &0&             &-1.&07855468(7)   &-0.&76054387(7) \\ 
\hline                                                                                                                                                                  
$(5,\,1)$    &-2.&70887493(5)   &2.&9502357(1)      &-1.&03742206(2)    &0&             &0&             &0&             &0&                &0&              \\ 
$(5,\,2)$    &-1.&35443746(2)   &1.&47511779(6)     &-0.&51871103(2)    &0&             &0&             &0&             &0&                &0&              \\ 
$(5,\,3)$    &0&                &0&                 &0&                 &0&             &0&             &0&             &0&                &0&              \\ 
$(5,\,4)$    &1.&1783609(6)     &-0.&49170598(2)    &-0.&15899133(2)    &0.&1590523(6)  &1/&3           &0&             &-0.&98220341(2)   &-0.&06601462(2) \\  
$(5,\,5)$    &2.&297078(2)      &-8.&35900166(4)    &-2.&70285255(5)    &10.&134698(2)  &17/&3          &-8/&3          &-3.&06215787(3)   &0.&83396857(3)  \\   
\hline
\hline
\multicolumn{1}{c}{$\phantom{(l,\,m)}$} &
\multicolumn{2}{c}{$\phantom{123456789012345}$} &
\multicolumn{2}{c}{$\phantom{123456789012345}$} &
\multicolumn{2}{c}{$\phantom{12345678901234}$} &
\multicolumn{2}{c}{$\phantom{123456789012345}$} &
\multicolumn{2}{c}{$\phantom{123456}$} &
\multicolumn{2}{c}{$\phantom{123456}$} &
\multicolumn{2}{c}{$\phantom{1234567890123}$} &
\multicolumn{2}{c}{$\phantom{1234567890123}$} \\
\end{tabular}
\end{minipage}
\end{center}
\vspace{-0.9cm}
\caption{{\sl The coefficients $ d_{l,m}^{+(0,1)} -  d_{l,m}^{+(0,6)}$ and 
$ d_{l,m}^{+(2,1)} -  d_{l,m}^{+(2,2)}$. }}
\label{4f1}
}}
\end{table}

\begin{table}[!h]
{\small{
\begin{center}
\begin{minipage}{24.5cm}
\begin{tabular}{|l|r@{}l|r@{}l|r@{}l|r@{}l|r@{}l|r@{}l|r@{}l|r@{}l|}
\multicolumn{17}{c}{\Red{Parity Conserving, Flavour Exchange Symmetry Plus }}\\
\hline
\hline
\multicolumn{1}{|c|}{$(l,\,m)$}&
\multicolumn{2}{c|}{$ d_{l,m}^{{+(2,3)}^{\phantom{A}}}$} &
\multicolumn{2}{c|}{$ d_{l,m}^{{+(2,4)}^{\phantom{A}}}$} &
\multicolumn{2}{c|}{$ d_{l,m}^{{+(2,5)}^{\phantom{A}}}$} &
\multicolumn{2}{c|}{$ d_{l,m}^{{+(2,6)}^{\phantom{A}}}$} &
\multicolumn{2}{c|}{$ d_{l,m}^{{+(2,7)}^{\phantom{A}}}$} &
\multicolumn{2}{c|}{$ d_{l,m}^{{+(2,8)}^{\phantom{A}}}$} &
\multicolumn{2}{c|}{$ d_{l,m}^{{+(2,9)}^{\phantom{A}}}$} &
\multicolumn{2}{c|}{$ d_{l,m}^{{+(2,10)}^{\phantom{A}}}$}\\
\hline
\hline
$(1,\,1)$    &2.&642227(4)      &1.&7473718(2)      &0.&08346609(1)     &-3.&289500(4)     &-19/&18     &-1/&2      &0&         &25/&24    \\  
$(1,\,2)$    &-1.&0988822(5)    &2.&5685206(3)      &-0.&9493554(4)     &0.&9569921(2)     &0&          &-1&        &1&         &-1/&3     \\ 
$(1,\,3)$    &0.&13736030(9)    &-0.&32106507(4)    &0.&11866942(8)     &-0.&11962402(4)   &0&          &1/&8       &-1/&8      &1/&24     \\ 
$(1,\,4)$    &0.&35245840(2)    &-0.&399645902(8)   &0.&06172133(2)     &0.&0040994899(4)  &-1/&4       &3/&8       &-1/&4      &-1/&24    \\ 
$(1,\,5)$    &1.&05737519(6)    &-1.&19893771(2)    &0.&18516399(5)     &0.&012298470(1)   &-3/&4       &9/&8       &-3/&4      &-1/&8     \\ 
\hline                                                                                                                                                 
$(2,\,1)$    &-1.&5109631(5)    &3.&5317158(3)      &-1.&3053636(4)     &1.&3158642(3)     &0&          &-11/&8     &11/&8      &-11/&24   \\ 
$(2,\,2)$    &0.&789419(4)      &1.&84106755(5)     &-0.&37712545(3)    &-2.&739903(4)     &1/&18       &-11/&8     &3/&8       &41/&48    \\  
$(2,\,3)$    &-1.&6447718(6)    &0.&199822951(3)    &-0.&065051477(5)   &-0.&0214962(6)    &11/&24      &-3/&16     &1/&16      &-1/&24    \\ 
$(2,\,4)$    &-0.&06868015(3)   &0.&16053253(1)     &-0.&01966766(2)    &0.&05981201(1)    &0&          &-1/&16     &0&         &-1/&48    \\ 
$(2,\,5)$    &0.&20604045(9)    &-0.&48159760(4)    &0.&05900298(7)     &-0.&17943602(4)   &0&          &3/&16      &0&         &1/&16     \\ 
\hline                                                                                                                                                 
$(3,\,1)$    &-0.&2747206(1)    &0.&64213013(6)     &-0.&2373388(1)   &0.&23924803(5)    &0&          &-1/&4      &1/&4       &-1/&12    \\ 
$(3,\,2)$    &-7.&416224(3)     &4.&6520712(2)      &-0.&18110545(2)    &0.&870210(3)      &5/&3        &-9/&4      &1/&4       &-13/&24   \\ 
$(3,\,3)$    &2.&045124(4)      &-3.&9381018(3)     &-0.&49577614(3)    &-4.&174196(4)     &11/&36      &7/&8       &3/&8       &17/&12    \\  
$(3,\,4)$    &1.&785683(1)      &-4.&1738459(5)     &0.&5113591(7)      &-1.&5551122(4)    &0&          &13/&8      &0&         &13/&24    \\ 
$(3,\,5)$    &1.&2362427(5)     &-2.&8895856(3)     &0.&3540179(4)      &-1.&0766161(2)    &0&          &9/&8       &0&         &3/&8      \\ 
\hline                                                                                                                                                 
$(4,\,1)$    &1.&40983359(8)    &-1.&59858361(4)    &0.&24688532(7)     &0.&016397959(2)   &-1&         &3/&2       &-1&        &-1/&6     \\ 
$(4,\,2)$    &-1.&9230442(7)    &4.&4949109(3)      &-0.&5506945(6)     &1.&6747362(3)     &0&          &-7/&4      &0&         &-7/&12    \\ 
$(4,\,3)$    &2.&197764(1)      &-5.&1370411(5)     &0.&6293651(7)      &-1.&9139843(4)    &0&          &2&         &0&         &2/&3      \\ 
$(4,\,4)$    &-0.&286209(4)     &-2.&8633160(3)     &-0.&023092586(8)   &-4.&273362(4)     &1/&18       &5/&4       &0&         &59/&48    \\  
$(4,\,5)$    &4.&602710(3)      &-3.&0948716(3)     &-0.&05164224(1)    &-0.&928529(3)     &-7/&9       &5/&4       &0&         &9/&16     \\ 
\hline                                                                                                                                                 
$(5,\,1)$    &1.&40983359(8)    &-1.&59858361(3)    &0.&24688532(6)     &0.&016397959(2)   &-1&         &3/&2       &-1&        &-1/&6     \\ 
$(5,\,2)$    &0.&2747206(1)     &-0.&64213013(6)    &0.&0786706(1)    &-0.&23924803(5)   &0&          &1/&4       &0&         &1/&12     \\ 
$(5,\,3)$    &0&                &0&                 &0&                 &0&                &0&          &0&         &0&         &0&        \\ 
$(5,\,4)$    &1.&255191(1)      &0.&2526361(1)    &0.&009152736(3)    &0.&009222(1)      &-17/&54     &-1/&12     &0&         &1/&16     \\  
$(5,\,5)$    &0.&828945(4)      &4.&0632560(3)      &0.&12704697(1)     &-2.&661259(4)     &-23/&27     &-17/&12    &0&         &35/&48    \\   
\hline
\hline
\multicolumn{1}{c}{$\phantom{(l,\,m)}$} &
\multicolumn{2}{c}{$\phantom{123456789012345}$} &
\multicolumn{2}{c}{$\phantom{123456789012345}$} &
\multicolumn{2}{c}{$\phantom{12345678901234}$} &
\multicolumn{2}{c}{$\phantom{123456789012345}$} &
\multicolumn{2}{c}{$\phantom{123456}$} &
\multicolumn{2}{c}{$\phantom{123456}$} &
\multicolumn{2}{c}{$\phantom{1234567890123}$} &
\multicolumn{2}{c}{$\phantom{1234567890123}$} \\
\end{tabular}
\end{minipage}
\end{center}
\vspace{-0.9cm}
\caption{{\sl The coefficients $ d_{l,m}^{+(2,3)} -  d_{l,m}^{+(2,10)}$. }}
\label{4f2}
}}
\end{table}

\begin{table}[!h]
{\small{
\begin{center}
\begin{minipage}{24.5cm}
\begin{tabular}{|l|r@{}l|r@{}l|r@{}l|r@{}l|r@{}l|r@{}l|r@{}l|r@{}l|}
\multicolumn{17}{c}{\Red{Parity Conserving, Flavour Exchange Symmetry Minus }}\\
\hline
\hline
\multicolumn{1}{|c|}{$(l,\,m)$}&
\multicolumn{2}{c|}{$ d_{l,m}^{{-(0,1)}^{\phantom{A}}}$} &
\multicolumn{2}{c|}{$ d_{l,m}^{{-(0,2)}^{\phantom{A}}}$} &
\multicolumn{2}{c|}{$ d_{l,m}^{{-(0,3)}^{\phantom{A}}}$} &
\multicolumn{2}{c|}{$ d_{l,m}^{{-(0,4)}^{\phantom{A}}}$} &
\multicolumn{2}{c|}{$ d_{l,m}^{{-(0,5)}^{\phantom{A}}}$} &
\multicolumn{2}{c|}{$ d_{l,m}^{{-(0,6)}^{\phantom{A}}}$} &
\multicolumn{2}{c|}{$ d_{l,m}^{{-(2,1)}^{\phantom{A}}}$} &
\multicolumn{2}{c|}{$ d_{l,m}^{{-(2,2)}^{\phantom{A}}}$}\\
\hline
\hline
$(1,\,1)$    &-2.&830716(3)     &5.&90047176(3)     &1.&90789592(4)    &9.&748573(3)     &-4&       &-8/&3      &2.&12575961(5)    &0.&46409210(5)  \\  
$(1,\,2)$    &5.&41774985(8)    &-5.&9004712(3)     &2.&0748441(1)     &0&               &0&        &0&         &0&                &0&              \\ 
$(1,\,3)$    &1.&35443746(2)    &-1.&47511779(6)    &0.&51871103(2)    &0&               &0&        &0&         &0&                &0&              \\ 
$(1,\,4)$    &1.&35443746(2)    &-1.&47511784(3)    &0.&518711032(7)   &0&               &0&        &0&         &0&                &0&              \\ 
$(1,\,5)$    &4.&06331239(5)    &-4.&42535353(9)    &1.&55613310(2)    &0&               &0&        &0&         &0&                &0&              \\ 
\hline                                                                                                                                                          
$(2,\,1)$    &3.&38609366(9)    &-3.&6877945(3)     &1.&2967776(1)     &0&               &0&        &0&         &0&                &0&              \\ 
$(2,\,2)$    &0.&267862(2)      &0.&73755883(5)     &-1.&25504053(1)   &10.&816593(2)    &1&        &-8/&3      &-0.&87361423(2)   &0.&57016434(2)  \\  
$(2,\,3)$    &1.&5317566(4)     &-0.&368779461(8)   &0.&129677758(2)   &0.&1931470(4)    &0&        &0&         &-0.&89270365(2)   &-0.&08962151(2) \\ 
$(2,\,4)$    &0.&338609366(5)   &-0.&36877945(2)    &0.&129677758(6)   &0&               &0&        &0&         &0&                &0&              \\ 
$(2,\,5)$    &-1.&01582810(1)   &1.&10633834(5)     &-0.&38903328(2)   &0&               &0&        &0&         &0&                &0&              \\ 
\hline                                                                                                                                                          
$(3,\,1)$    &1.&35443746(2)    &-1.&47511779(6)    &0.&51871103(2)    &0&               &0&        &0&         &0&                &0&              \\ 
$(3,\,2)$    &12.&324653(1)     &7.&37558979(4)     &3.&38055490(3)    &3.&772588(1)     &-6&       &0&         &-1.&70275904(9)   &-0.&92098603(9) \\ 
$(3,\,3)$    &9.&564301(2)      &14.&01362029(6)    &3.&03772527(5)    &15.&316593(2)    &-8&       &-8/&3      &1.&92846910(2)    &-0.&27358566(2) \\  
$(3,\,4)$    &-12.&8671559(2)   &14.&0136190(6)     &-4.&9277548(2)    &0&               &0&        &0&         &0&                &0&              \\ 
$(3,\,5)$    &6.&09496858(8)    &-6.&6380301(3)     &2.&3341997(1)     &0&               &0&        &0&         &0&                &0&              \\ 
\hline                                                                                                                                                          
$(4,\,1)$    &5.&41774986(5)    &-5.&9004714(1)     &2.&07484413(3)    &0&               &0&        &0&         &0&                &0&              \\ 
$(4,\,2)$    &-6.&7721873(1)    &7.&3755889(4)      &-2.&5935552(1)    &0&               &0&        &0&         &0&                &0&              \\ 
$(4,\,3)$    &-10.&8354997(2)   &11.&8009423(6)     &-4.&1496883(2)    &0&               &0&        &0&         &0&                &0&              \\ 
$(4,\,4)$    &12.&922182(2)     &16.&22629734(4)    &5.&24671376(5)    &15.&816593(2)    &-11&      &-8/&3      &1.&84126332(3)    &-0.&57566914(3) \\  
$(4,\,5)$    &-9.&555272(2)     &-1.&47511794(3)    &-0.&47697397(5)   &-3.&068020(2)    &1&        &0&         &2.&15255184(7)    &0.&47726124(7)  \\ 
\hline                                                                                                                                                          
$(5,\,1)$    &5.&41774986(6)    &-5.&9004714(1)     &2.&07484413(3)    &0&               &0&        &0&         &0&                &0&              \\ 
$(5,\,2)$    &-1.&35443746(2)   &1.&47511779(6)     &-0.&51871103(2)   &0&               &0&        &0&         &0&                &0&              \\ 
$(5,\,3)$    &0&                &0&                 &0&                &0&               &0&        &0&         &0&                &0&              \\ 
$(5,\,4)$    &-1.&1192154(6)    &2.&45852990(2)     &0.&79495663(2)    &-0.&0226732(6)   &-5/&3     &0&         &1.&34020247(2)    &-0.&02841292(2) \\  
$(5,\,5)$    &-3.&777376(2)     &0.&49170598(4)     &0.&15899132(5)    &8.&771247(2)     &-1/&3     &-8/&3      &0.&16287196(3)    &0.&68000502(3)  \\   
\hline
\hline
\multicolumn{1}{c}{$\phantom{(l,\,m)}$} &
\multicolumn{2}{c}{$\phantom{123456789012345}$} &
\multicolumn{2}{c}{$\phantom{123456789012345}$} &
\multicolumn{2}{c}{$\phantom{12345678901234}$} &
\multicolumn{2}{c}{$\phantom{123456789012345}$} &
\multicolumn{2}{c}{$\phantom{123456}$} &
\multicolumn{2}{c}{$\phantom{123456}$} &
\multicolumn{2}{c}{$\phantom{1234567890123}$} &
\multicolumn{2}{c}{$\phantom{1234567890123}$} \\
\end{tabular}
\end{minipage}
\end{center}
\vspace{-0.9cm}
\caption{{\sl The coefficients $ d_{l,m}^{-(0,1)} -  d_{l,m}^{-(0,6)}$ and 
$ d_{l,m}^{-(2,1)} -  d_{l,m}^{-(2,2)}$. }}
\label{4f3}
}}
\end{table}

\begin{table}[!h]
{\small{
\begin{center}
\begin{minipage}{24.5cm}
\begin{tabular}{|l|r@{}l|r@{}l|r@{}l|r@{}l|r@{}l|r@{}l|r@{}l|r@{}l|}
\multicolumn{17}{c}{\Red{Parity Conserving, Flavour Exchange Symmetry Minus }}\\
\hline
\hline
\multicolumn{1}{|c|}{$(l,\,m)$}&
\multicolumn{2}{c|}{$ d_{l,m}^{{-(2,3)}^{\phantom{A}}}$} &
\multicolumn{2}{c|}{$ d_{l,m}^{{-(2,4)}^{\phantom{A}}}$} &
\multicolumn{2}{c|}{$ d_{l,m}^{{-(2,5)}^{\phantom{A}}}$} &
\multicolumn{2}{c|}{$ d_{l,m}^{{-(2,6)}^{\phantom{A}}}$} &
\multicolumn{2}{c|}{$ d_{l,m}^{{-(2,7)}^{\phantom{A}}}$} &
\multicolumn{2}{c|}{$ d_{l,m}^{{-(2,8)}^{\phantom{A}}}$} &
\multicolumn{2}{c|}{$ d_{l,m}^{{-(2,9)}^{\phantom{A}}}$} &
\multicolumn{2}{c|}{$ d_{l,m}^{{-(2,10)}^{\phantom{A}}}$}\\
\hline
\hline
$(1,\,1)$   &3.&611581(4)       &-3.&4947437(3)    &-0.&16693217(1)    &-2.&043017(4)      &-5/&9        &1&            &0&        &5/&12     \\  
$(1,\,2)$   &-1.&0988822(5)     &2.&5685206(3)     &-0.&9493554(4)     &0.&9569921(2)      &0&           &-1&           &1&        &-1/&3     \\ 
$(1,\,3)$   &-0.&2747206(1)     &0.&64213013(6)    &-0.&2373388(1)   &0.&23924803(5)     &0&           &-1/&4         &1/&4      &-1/&12    \\ 
$(1,\,4)$   &-0.&70491680(2)    &0.&79929180(1)    &-0.&12344266(2)    &-0.&0081989797(6)  &1/&2         &-3/&4         &1/&2      &1/&12     \\ 
$(1,\,5)$   &-2.&11475039(7)    &2.&39787541(3)    &-0.&37032798(6)    &-0.&024596939(2)   &3/&2         &-9/&4         &3/&2      &1/&4      \\ 
\hline                                                                                                                                                    
$(2,\,1)$   &-0.&6868013(5)     &1.&6053254(3)     &-0.&5933471(4)     &0.&5981201(2)      &0&           &-5/&8         &5/&8      &-5/&24    \\ 
$(2,\,2)$   &2.&904169(4)       &-0.&55680786(5)   &0.&40349227(3)     &-2.&715306(4)      &-13/&9       &7/&8          &-3/&8     &29/&48    \\  
$(2,\,3)$   &1.&2923134(6)      &0.&199822951(3)   &-0.&065051477(5)   &0.&0173967(6)      &-5/&24       &-3/&16        &1/&16     &1/&12     \\ 
$(2,\,4)$   &-0.&06868015(3)    &0.&16053253(1)    &-0.&01966766(2)    &0.&05981201(1)     &0&           &-1/&16        &0&        &-1/&48    \\ 
$(2,\,5)$   &0.&20604045(9)     &-0.&48159760(4)   &0.&05900298(7)     &-0.&17943602(4)    &0&           &3/&16         &0&        &1/&16     \\ 
\hline                                                                                                                                                    
$(3,\,1)$   &-0.&2747206(1)     &0.&64213013(6)    &-0.&2373388(1)   &0.&23924803(5)     &0&           &-1/&4         &1/&4      &-1/&12    \\ 
$(3,\,2)$   &6.&006391(3)       &-3.&0534875(2)    &-0.&33930637(2)    &-0.&886608(3)      &-2/&3        &3/&4          &1/&4      &17/&24    \\ 
$(3,\,3)$   &4.&159875(4)       &-6.&3359772(3)    &0.&28484158(3)     &-4.&149599(4)      &-43/&36      &25/&8         &-3/&8     &7/&6      \\  
$(3,\,4)$   &2.&609845(1)       &-6.&1002363(5)    &0.&7473711(7)      &-2.&2728563(4)     &0&           &19/&8         &0&        &19/&24    \\ 
$(3,\,5)$   &-1.&2362427(5)     &2.&8895856(3)     &-0.&3540179(4)     &1.&0766161(2)      &0&           &-9/&8         &0&        &-3/&8     \\ 
\hline                                                                                                                                                    
$(4,\,1)$   &-2.&81966719(8)    &3.&19716722(4)    &-0.&49377065(7)    &-0.&032795919(2)   &2&           &-3&           &2&        &1/&3      \\ 
$(4,\,2)$   &1.&3736030(7)      &-3.&2106507(3)    &0.&3933532(6)      &-1.&1962402(3)     &0&           &5/&4          &0&        &5/&12     \\ 
$(4,\,3)$   &2.&197764(1)       &-5.&1370411(5)    &0.&6293651(7)      &-1.&9139843(4)     &0&           &2&            &0&        &2/&3      \\ 
$(4,\,4)$   &6.&491207(4)       &-7.&4107631(3)    &-0.&187841964(8)   &-4.&050432(4)      &-17/&18      &11/&4         &0&        &65/&48    \\  
$(4,\,5)$   &-2.&042489(3)      &0.&0632400(3)     &-0.&05819067(1)    &1.&051220(3)       &5/&9         &-1/&4         &0&        &-9/&16    \\ 
\hline                                                                                                                                                    
$(5,\,1)$   &-2.&8196672(1)   &3.&19716722(4)    &-0.&49377065(8)    &-0.&032795919(2)   &2&           &-3&           &2&        &1/&3      \\ 
$(5,\,2)$   &0.&2747206(1)      &-0.&64213013(6)   &0.&0786706(1)    &-0.&23924803(5)    &0&           &1/&4          &0&        &1/&12     \\ 
$(5,\,3)$   &0&                 &0&                &0&                 &0&                 &0&           &0&            &0&        &0&        \\ 
$(5,\,4)$   &-0.&401784(1)      &-1.&2631800(1)    &-0.&045763706(3)   &0.&031675(1)       &13/&54       &5/&12         &0&        &-1/&16    \\  
$(5,\,5)$   &3.&734320(4)       &-0.&9473033(3)    &-0.&09480166(1)    &-1.&755961(4)      &-23/&27      &1/&12         &0&        &17/&48    \\   
\hline
\hline
\multicolumn{1}{c}{$\phantom{(l,\,m)}$} &
\multicolumn{2}{c}{$\phantom{123456789012345}$} &
\multicolumn{2}{c}{$\phantom{123456789012345}$} &
\multicolumn{2}{c}{$\phantom{12345678901234}$} &
\multicolumn{2}{c}{$\phantom{123456789012345}$} &
\multicolumn{2}{c}{$\phantom{123456}$} &
\multicolumn{2}{c}{$\phantom{123456}$} &
\multicolumn{2}{c}{$\phantom{1234567890123}$} &
\multicolumn{2}{c}{$\phantom{1234567890123}$} \\
\end{tabular}
\end{minipage}
\end{center}
\vspace{-0.9cm}
\caption{{\sl The coefficients $ d_{l,m}^{-(2,3)} -  d_{l,m}^{-(2,10)}$. }}
\label{4f4}
}}
\end{table}

\begin{table}[!h]
{\small{
\begin{center}
\begin{minipage}{24.5cm}
\begin{tabular}{|l|r@{}l|r@{}l|r@{}l|r@{}l|r@{}l|r@{}l|r@{}l|r@{}l|}
\multicolumn{17}{c}{\Red{Parity Violating, Flavour Exchange Symmetry Plus }}\\
\hline
\hline
\multicolumn{1}{|c|}{$(l,\,m)$}&
\multicolumn{2}{c|}{$ \delta_{l,m}^{{+(0,1)}^{\phantom{A}}}$} &
\multicolumn{2}{c|}{$ \delta_{l,m}^{{+(0,2)}^{\phantom{A}}}$} &
\multicolumn{2}{c|}{$ \delta_{l,m}^{{+(0,3)}^{\phantom{A}}}$} &
\multicolumn{2}{c|}{$ \delta_{l,m}^{{+(0,4)}^{\phantom{A}}}$} &
\multicolumn{2}{c|}{$ \delta_{l,m}^{{+(0,5)}^{\phantom{A}}}$} &
\multicolumn{2}{c|}{$ \delta_{l,m}^{{+(0,6)}^{\phantom{A}}}$} &
\multicolumn{2}{c|}{$ \delta_{l,m}^{{+(2,1)}^{\phantom{A}}}$} &
\multicolumn{2}{c|}{$ \delta_{l,m}^{{+(2,2)}^{\phantom{A}}}$}\\
\hline
\hline
$(1,\,1)$     &7.&607190(2)     &-2.&95023588(3)    &-0.&95394796(3)    &12.&293750(2)   &2&        &-8/&3     &-2.&79899092(5)   &0.&39295395(5)  \\  
\hline                                                                                                                                                         
$(2,\,2)$     &2.&299519(2)     &-1.&475117940(5)   &-0.&476973978(5)   &10.&816593(2)   &1&        &-8/&3     &-0.&87361423(2)   &0.&57016434(2)  \\  
$(2,\,3)$     &-1.&1931472(4)   &0&                 &0&                 &-0.&1931470(4)  &0&        &0&        &0.&89270365(2)    &0.&08962151(2)  \\ 
\hline                                                                                                                                                         
$(3,\,2)$     &-10.&970215(1)   &-8.&85070764(3)    &-2.&86184387(3)    &-3.&772588(1)   &6&        &0&        &1.&70275904(9)    &0.&92098603(9)  \\ 
$(3,\,3)$     &11.&595958(2)    &11.&80094352(4)    &3.&81579182(4)     &15.&316593(2)   &-8&       &-8/&3     &1.&92846910(2)    &-0.&27358566(2) \\  
\hline                                                                                                                                                         
$(4,\,4)$     &10.&269733(2)    &7.&37558970(4)     &2.&38486989(5)     &14.&816594(2)   &-5&       &-8/&3     &2.&01567488(3)    &0.&02849782(3)  \\  
$(4,\,5)$     &9.&732709(2)     &7.&37558970(3)     &2.&38486988(5)     &3.&477157(2)    &-5&       &0&        &-1.&07855468(7)   &-0.&76054387(7) \\ 
\hline                                                                                                                                                         
$(5,\,4)$     &1.&1783609(6)    &-0.&49170598(2)    &-0.&15899133(2)    &0.&1590523(6)   &1/&3      &0&        &-0.&98220341(2)   &-0.&06601462(2) \\  
$(5,\,5)$     &2.&297078(2)     &-8.&35900166(4)    &-2.&70285255(5)    &10.&134698(2)   &17/&3     &-8/&3     &-3.&06215787(3)   &0.&83396857(3)  \\   
\hline
\hline
\multicolumn{1}{c}{$\phantom{(l,\,m)}$} &
\multicolumn{2}{c}{$\phantom{123456789012345}$} &
\multicolumn{2}{c}{$\phantom{123456789012345}$} &
\multicolumn{2}{c}{$\phantom{12345678901234}$} &
\multicolumn{2}{c}{$\phantom{123456789012345}$} &
\multicolumn{2}{c}{$\phantom{123456}$} &
\multicolumn{2}{c}{$\phantom{123456}$} &
\multicolumn{2}{c}{$\phantom{1234567890123}$} &
\multicolumn{2}{c}{$\phantom{1234567890123}$} \\
\end{tabular}
\end{minipage}
\end{center}
\vspace{-0.9cm}
\caption{{\sl The coefficients $ \delta_{l,m}^{+(0,1)} -  \delta_{l,m}^{+(0,6)}$ and 
$ \delta_{l,m}^{+(2,1)} -  \delta_{l,m}^{+(2,2)}$. }}
\label{4f5}
}}
%
%
{\small{
\begin{center}
\begin{minipage}{24.5cm}
\begin{tabular}{|l|r@{}l|r@{}l|r@{}l|r@{}l|r@{}l|r@{}l|r@{}l|r@{}l|}
\multicolumn{17}{c}{\Red{Parity Violating, Flavour Exchange Symmetry Plus }}\\
\hline
\hline
\multicolumn{1}{|c|}{$(l,\,m)$}&
\multicolumn{2}{c|}{$ \delta_{l,m}^{{+(2,3)}^{\phantom{A}}}$} &
\multicolumn{2}{c|}{$ \delta_{l,m}^{{+(2,4)}^{\phantom{A}}}$} &
\multicolumn{2}{c|}{$ \delta_{l,m}^{{+(2,5)}^{\phantom{A}}}$} &
\multicolumn{2}{c|}{$ \delta_{l,m}^{{+(2,6)}^{\phantom{A}}}$} &
\multicolumn{2}{c|}{$ \delta_{l,m}^{{+(2,7)}^{\phantom{A}}}$} &
\multicolumn{2}{c|}{$ \delta_{l,m}^{{+(2,8)}^{\phantom{A}}}$} &
\multicolumn{2}{c|}{$ \delta_{l,m}^{{+(2,9)}^{\phantom{A}}}$} &
\multicolumn{2}{c|}{$ \delta_{l,m}^{{+(2,10)}^{\phantom{A}}}$}\\
\hline
\hline
$(1,\,1)$    &2.&642227(4)     &1.&7473718(2)     &0.&08346609(1)     &-3.&289500(4)   &-19/&18        &-1/&2       &$\,\,\,\,\,$0&      &25/&24  \\  
\hline                                                                                                                                            
$(2,\,2)$    &1.&846794(4)     &0.&64212984(5)    &0.&0131834100(7)   &-2.&727604(4)   &-25/&36        &-1/&4        &$\,\,\,\,\,$0&      &35/&48  \\  
$(2,\,3)$    &-1.&4685426(6)   &0&                &0&                 &-0.&0194464(6)  &1/&3           &0&          &$\,\,\,\,\,$0&      &-1/&16  \\ 
\hline                                                                                                                                            
$(3,\,2)$    &-6.&711307(3)    &3.&8527794(2)     &0.&079100460(4)    &0.&878409(3)    &7/&6           &-3/&2       &$\,\,\,\,\,$0&      &-5/&8   \\ 
$(3,\,3)$    &3.&102499(4)     &-5.&1370395(3)    &-0.&105467280(7)   &-4.&161897(4)   &-4/&9          &2&          &$\,\,\,\,\,$0&      &31/&24  \\  
\hline                                                                                                                                            
$(4,\,4)$    &-0.&286209(4)    &-2.&8633160(3)    &-0.&023092586(8)   &-4.&273362(4)   &1/&18          &5/&4        &$\,\,\,\,\,$0&      &59/&48  \\  
$(4,\,5)$    &4.&602710(3)     &-3.&0948716(3)    &-0.&05164224(1)    &-0.&928529(3)   &-7/&9          &5/&4        &$\,\,\,\,\,$0&      &9/&16   \\ 
\hline                                                                                                                                            
$(5,\,4)$    &1.&255191(1)     &0.&2526361(1)   &0.&009152736(3)    &0.&009222(1)    &-17/&54        &-1/&12        &$\,\,\,\,\,$0&      &1/&16   \\  
$(5,\,5)$    &0.&828945(4)     &4.&0632560(3)     &0.&12704697(1)     &-2.&661259(4)   &-23/&27        &-17/&12     &$\,\,\,\,\,$0&      &35/&48  \\   
\hline
\hline
\multicolumn{1}{c}{$\phantom{(l,\,m)}$} &
\multicolumn{2}{c}{$\phantom{123456789012345}$} &
\multicolumn{2}{c}{$\phantom{123456789012345}$} &
\multicolumn{2}{c}{$\phantom{12345678901234}$} &
\multicolumn{2}{c}{$\phantom{12345678901234}$} &
\multicolumn{2}{c}{$\phantom{123456}$} &
\multicolumn{2}{c}{$\phantom{123456}$} &
\multicolumn{2}{c}{$\phantom{1234567890123}$} &
\multicolumn{2}{c}{$\phantom{1234567890123}$} \\
\end{tabular}
\end{minipage}
\end{center}
\vspace{-0.9cm}
\caption{{\sl The coefficients $ \delta_{l,m}^{+(2,3)} -  \delta_{l,m}^{+(2,10)}$. }}
\label{4f6}
}}
\end{table}

\begin{table}[!h]
{\small{
\begin{center}
\begin{minipage}{24.5cm}
\begin{tabular}{|l|r@{}l|r@{}l|r@{}l|r@{}l|r@{}l|r@{}l|r@{}l|r@{}l|}
\multicolumn{17}{c}{\Red{Parity Violating, Flavour Exchange Symmetry Minus }}\\
\hline
\hline
\multicolumn{1}{|c|}{$(l,\,m)$}&
\multicolumn{2}{c|}{$ \delta_{l,m}^{{-(0,1)}^{\phantom{A}}}$} &
\multicolumn{2}{c|}{$ \delta_{l,m}^{{-(0,2)}^{\phantom{A}}}$} &
\multicolumn{2}{c|}{$ \delta_{l,m}^{{-(0,3)}^{\phantom{A}}}$} &
\multicolumn{2}{c|}{$ \delta_{l,m}^{{-(0,4)}^{\phantom{A}}}$} &
\multicolumn{2}{c|}{$ \delta_{l,m}^{{-(0,5)}^{\phantom{A}}}$} &
\multicolumn{2}{c|}{$ \delta_{l,m}^{{-(0,6)}^{\phantom{A}}}$} &
\multicolumn{2}{c|}{$ \delta_{l,m}^{{-(2,1)}^{\phantom{A}}}$} &
\multicolumn{2}{c|}{$ \delta_{l,m}^{{-(2,2)}^{\phantom{A}}}$}\\
\hline
\hline
$(1,\,1)$    &-2.&830716(3)    &5.&90047176(3)     &1.&90789592(4)     &9.&748573(3)    &-4&       &-8/&3      &2.&12575961(5)    &0.&46409210(5)  \\  
\hline                                                                                                                                                         
$(2,\,2)$    &2.&299519(2)     &-1.&475117940(5)   &-0.&476973978(5)   &10.&816593(2)   &1&        &-8/&3      &-0.&87361423(2)   &0.&57016434(2)  \\  
$(2,\,3)$    &1.&1931472(4)    &0&                 &0&                 &0.&1931470(4)   &0&        &0&         &-0.&89270365(2)   &-0.&08962151(2) \\ 
\hline                                                                                                                                                         
$(3,\,2)$    &10.&970215(1)    &8.&85070764(3)     &2.&86184387(3)     &3.&772588(1)    &-6&       &0&         &-1.&70275904(9)   &-0.&92098603(9) \\ 
$(3,\,3)$    &11.&595958(2)    &11.&80094352(4)    &3.&81579182(4)     &15.&316593(2)   &-8&       &-8/&3      &1.&92846910(2)    &-0.&27358566(2) \\  
\hline                                                                                                                                                         
$(4,\,4)$    &12.&922182(2)    &16.&22629734(4)    &5.&24671376(5)     &15.&816593(2)   &-11&      &-8/&3      &1.&84126332(3)    &-0.&57566914(3) \\  
$(4,\,5)$    &-9.&555272(2)    &-1.&47511794(3)    &-0.&47697397(5)    &-3.&068020(2)   &1&        &0&         &2.&15255184(7)    &0.&47726124(7)  \\ 
\hline                                                                                                                                                         
$(5,\,4)$    &-1.&1192154(6)   &2.&45852990(2)     &0.&79495663(2)     &-0.&0226732(6)  &-5/&3     &0&         &1.&34020247(2)    &-0.&02841292(2) \\  
$(5,\,5)$    &-3.&777376(2)    &0.&49170598(4)     &0.&15899132(5)     &8.&771247(2)    &-1/&3     &-8/&3      &0.&16287196(3)    &0.&68000502(3)  \\   
\hline
\hline
\multicolumn{1}{c}{$\phantom{(l,\,m)}$} &
\multicolumn{2}{c}{$\phantom{123456789012345}$} &
\multicolumn{2}{c}{$\phantom{123456789012345}$} &
\multicolumn{2}{c}{$\phantom{12345678901234}$} &
\multicolumn{2}{c}{$\phantom{123456789012345}$} &
\multicolumn{2}{c}{$\phantom{123456}$} &
\multicolumn{2}{c}{$\phantom{123456}$} &
\multicolumn{2}{c}{$\phantom{1234567890123}$} &
\multicolumn{2}{c}{$\phantom{1234567890123}$} \\
\end{tabular}
\end{minipage}
\end{center}
\vspace{-0.9cm}
\caption{{\sl The coefficients $ \delta_{l,m}^{-(0,1)} -  \delta_{l,m}^{-(0,6)}$ and 
$ \delta_{l,m}^{-(2,1)} -  \delta_{l,m}^{-(2,2)}$. }}
\label{4f7}
}}
%
{\small{
\begin{center}
\begin{minipage}{24.5cm}
\begin{tabular}{|l|r@{}l|r@{}l|r@{}l|r@{}l|r@{}l|r@{}l|r@{}l|r@{}l|}
\multicolumn{17}{c}{\Red{Parity Violating, Flavour Exchange Symmetry Minus }}\\
\hline
\hline
\multicolumn{1}{|c|}{$(l,\,m)$}&
\multicolumn{2}{c|}{$ \delta_{l,m}^{{-(2,3)}^{\phantom{A}}}$} &
\multicolumn{2}{c|}{$ \delta_{l,m}^{{-(2,4)}^{\phantom{A}}}$} &
\multicolumn{2}{c|}{$ \delta_{l,m}^{{-(2,5)}^{\phantom{A}}}$} &
\multicolumn{2}{c|}{$ \delta_{l,m}^{{-(2,6)}^{\phantom{A}}}$} &
\multicolumn{2}{c|}{$ \delta_{l,m}^{{-(2,7)}^{\phantom{A}}}$} &
\multicolumn{2}{c|}{$ \delta_{l,m}^{{-(2,8)}^{\phantom{A}}}$} &
\multicolumn{2}{c|}{$ \delta_{l,m}^{{-(2,9)}^{\phantom{A}}}$} &
\multicolumn{2}{c|}{$ \delta_{l,m}^{{-(2,10)}^{\phantom{A}}}$}\\
\hline
\hline
$(1,\,1)$    &3.&611581(4)    &-3.&4947437(3)   &-0.&16693217(1)    &-2.&043017(4)   &-5/&9          &1&        &$\,\,\,\,\,$0&        &5/&12   \\  
\hline                                                                                                                                                      
$(2,\,2)$    &1.&846794(4)    &0.&64212984(5)   &0.&0131834100(7)   &-2.&727604(4)   &-25/&36        &-1/&4     &$\,\,\,\,\,$0&        &35/&48  \\  
$(2,\,3)$    &1.&4685426(6)   &0&               &0&                 &0.&0194464(6)   &-1/&3          &0&        &$\,\,\,\,\,$0&        &1/&16   \\ 
\hline                                                                                                                                                      
$(3,\,2)$    &6.&711307(3)    &-3.&8527794(2)   &-0.&079100460(4)   &-0.&878409(3)   &-7/&6          &3/&2      &$\,\,\,\,\,$0&        &5/&8    \\ 
$(3,\,3)$    &3.&102499(4)    &-5.&1370395(3)   &-0.&105467280(7)   &-4.&161897(4)   &-4/&9          &2&        &$\,\,\,\,\,$0&        &31/&24  \\  
\hline                                                                                                                                                      
$(4,\,4)$    &6.&491207(4)    &-7.&4107631(3)   &-0.&187841964(8)   &-4.&050432(4)   &-17/&18        &11/&4     &$\,\,\,\,\,$0&        &65/&48  \\  
$(4,\,5)$    &-2.&042489(3)   &0.&0632400(3)    &-0.&05819067(1)    &1.&051220(3)    &5/&9           &-1/&4     &$\,\,\,\,\,$0&        &-9/&16  \\ 
\hline                                                                                                                                                      
$(5,\,4)$    &-0.&401784(1)   &-1.&2631800(1)   &-0.&045763706(3)   &0.&031675(1)    &13/&54         &5/&12     &$\,\,\,\,\,$0&        &-1/&16  \\  
$(5,\,5)$    &3.&734320(4)    &-0.&9473033(3)   &-0.&09480166(1)    &-1.&755961(4)   &-23/&27        &1/&12     &$\,\,\,\,\,$0&        &17/&48  \\   
\hline
\hline
\multicolumn{1}{c}{$\phantom{(l,\,m)}$} &
\multicolumn{2}{c}{$\phantom{123456789012345}$} &
\multicolumn{2}{c}{$\phantom{123456789012345}$} &
\multicolumn{2}{c}{$\phantom{12345678901234}$} &
\multicolumn{2}{c}{$\phantom{12345678901234}$} &
\multicolumn{2}{c}{$\phantom{123456}$} &
\multicolumn{2}{c}{$\phantom{123456}$} &
\multicolumn{2}{c}{$\phantom{1234567890123}$} &
\multicolumn{2}{c}{$\phantom{1234567890123}$} \\
\end{tabular}
\end{minipage}
\end{center}
\vspace{-0.9cm}
\caption{{\sl The coefficients $ \delta_{l,m}^{-(2,3)} -  \delta_{l,m}^{-(2,10)}$. }}
\label{4f8}
}}
\end{table}
\end{turnpage}

\end{document}